\documentclass[11pt]{article}
\usepackage{jcappub}
\usepackage{bm}
\usepackage{color}
\usepackage{graphicx}
\usepackage{xcolor}
\usepackage{multirow}
\usepackage{epstopdf}
\usepackage{geometry} 
\usepackage{slashed}

\newcommand{\qmsq}{\qmsq}
\renewcommand{\qmsq}{\q^2}

\newcommand{\MG}{\:M_{\text{GUT}}}

\newcommand{\MEW}{\:M_{\text{EW}}}
\newcommand{\MR}{\:M_{R}}

\newcommand{\eq}[1]{Eq.~\eqref{#1}}

\newcommand{\Figref}[1]{Fig.~\ref{#1}}
\newcommand{\Tabref}[1]{Tab.~\ref{#1}}

\newcommand{\be}{\begin{equation}}
\newcommand{\ee}{\end{equation}}
\def\bea{\begin{eqnarray} }
\def\eea{ \end{eqnarray} } 
\newcommand{\een}{\end{subequations}}
\newcommand{\ben}{\begin{subequations}}
\newcommand{\beq}{\begin{eqalignno}}
\newcommand{\eeq}{\end{eqalignno}}

\newcommand{\lsim}{\mathrel{\mathop{\kern 0pt \rlap
      {\raise.2ex\hbox{$<$}}}\lower.9ex\hbox{\kern-.190em $ \sim$}}}
\newcommand{\gsim}{\mathrel{\mathop{\kern 0pt
      \rlap{\raise.2ex\hbox{$>$}}}\lower.9ex\hbox{\kern-.190em $\sim$}}}

\newcommand{\VectorTypefaceArrow}{
\let\oldvec\vec
\renewcommand{\vec}[1]{\oldvec{##1}} 
\newcommand{\uvec}[1]{\hat{##1}} 
}

\VectorTypefaceArrow

\usepackage{upgreek} 
\newcommand{\q}{{\widetilde{q}}}

\usepackage{mathtools}

\usepackage{booktabs}
\usepackage{xcolor}
\usepackage{arydshln}
\usepackage{cancel}

\allowdisplaybreaks

\author[a,b]{Anirban Biswas,}
\author[a,b]{Arpan Kar,}
\author[a,b]{Hyomin Kim,}
\author[a,b]{Stefano Scopel,}
\author[a,b,c]{Liliana Velasco-Sevilla,}
\affiliation[a]{Center for Quantum Spacetime, Sogang University, Seoul 121-742, South Korea}
\affiliation[b]{Department of Physics, Sogang University, Seoul 121-742, South Korea}
\affiliation[c]{Korea Institute for Advanced Study, Seoul 02455, South Korea}
\emailAdd{anirban.biswas.sinp@gmail.com}
\emailAdd{arpankarphys@gmail.com}
\emailAdd{hyomin1996@naver.com}
\emailAdd{scopel@sogang.ac.kr}
\emailAdd{liliana.velascosevilla@gmail.com}

\title{Improved White Dwarves Constraints on Inelastic Dark Matter and Left--Right Symmetric Models}

\abstract{Weakly Interacting Massive Particles (WIMPs) can be captured in compact stars such as white dwarves (WDs) if they are in a Dark Matter (DM)--rich environment, leading to an increase in the star luminosity through their annihilation process. N--body simulations suggest that the core of the Messier 4 globular cluster (where plenty of WDs are observed) is rich of Dark Matter. Assuming this is the case, we use a recent improvement in the calculation of the WD equation of state to show that when the WIMP interacts with the nuclear targets within the WD through inelastic scattering, and its mass exceeds a few tens GeV, the data on low--temperature large-mass WDs in M4 can probe values of the mass splitting as large as $\delta\lsim$ 40 MeV. Such value largely exceeds those ensuing from direct detection and from solar neutrino searches.  We apply such improved constraint to the specific DM scenario of a self--conjugate bi-doublet in the Left--Right Symmetric Model (LRSM), where the standard $SU(2)_L$ group with coupling $g_L$ is extended by an additional $SU(2)_R$ with coupling $g_R$ in order to explain maximal parity violation in Weak Interactions. We show that bounds from WDs significantly reduce the cosmologically viable parameter space of such scenario, in particular requiring $g_R>g_L$. For instance, for $g_R/g_L$ = 1.8 we find the two viable mass ranges 1.2 TeV $\lsim m_\chi\lsim$ 3 TeV and 5 TeV $\lsim m_\chi\lsim$ 10 TeV, when the charged $SU(2)_R$ gauge boson mass $M_{W_2}$ is lighter than $\simeq$ 12 TeV. We also discuss the ultraviolet completion of the LRSM model, when the latter is embedded in a Grand Unified Theory. We show that such low--energy parameter space and compatibility to proton-decay bounds require a non--trivial extension of the particle content of the minimal model. We provide a specific example where $M_{W_2}\lsim$ 10 TeV is achieved by extending the LRSM at high energy with color triplets that are singlets under all other groups, and $g_R/g_L>$1 is obtained by introducing $SU(2)_L$ triplets with no $SU(2)_R$ counterparts, i.e. by breaking the symmetry between the multiplets of $SU(2)_L$ and $SU(2)_R$.}

\begin{document}
\hspace*{87.0mm}{CQUeST-2022-0691, KIAS  P22038}\\
\maketitle
\section{Introduction}
\label{sec:introduction}

The discovery of the nature of Cold Dark Matter (CDM) is one of the most urgent pending issues of modern physics. Weakly Interacting Massive Particles (WIMPs) 
are the most popular CDM candidates: their mass $m_\chi$ falls in the GeV–TeV range and they are expected to have only weak-type interactions with ordinary matter.  Such small but non--vanishing interactions keep them in thermal equilibrium in the Early Universe, providing an expected thermal relic abundance in agreement with observation. They also provide promising signals to detect them: in direct detection (DD) WIMP scattering events off nuclear targets are searched for through the measurement of the ensuing nuclear recoils in low-background detectors 
(see for instance~\cite{DD_Schumann2019, Snowmass_Leane2022}); in accelerators WIMPs can be created from the scattering of Standard Model (SM) initial states (see for instance~\cite{collider_Arcadi2017, collider_Boveia2018, collider_Giagu2019}); finally, in indirect detection (ID) the products of the annihilations of DM particles in the halo of our Galaxy or that of dwarf galaxies are searched for both in ground--based and in satellite telescopes 
(see for example~\cite{ID_Gaskins2016, ID_Strigari2018, ID_Leane2020, ID_PerezdelosHeros2020}). 

Annihilation is enhanced wherever the WIMP density is large, so that a peculiar ID signal is provided by celestial bodies such as the Sun or the Earth, that, through the same WIMP--nucleus scattering processes that are searched for in DD, can capture WIMPs developing a high--density population of them in their interior~\cite{capture_Press_Spergel, capture_Gould_1987ApJ, capture_Gould_1988ApJ, gould_capture_1991, dm_cap_sun_PhysRevD.86.073002, dm_capture_sun_Catena2015, dm_capture_sun_Brenner2020}. The opacity of the celestial body plasma implies that among the WIMP annihilation products only neutrinos can escape from its surface, with energies that can reach about one third of the WIMP mass~\cite{cap_nu_sun_PhysRevLett.55.257, cap_nu_sun_HAGELIN1986, dm_cap_nu_SREDNICKI1987, Jungman:1994jr}. The detection of neutrinos coming out from the Sun or the Earth with energies ranging from the GeV to the TeV scale would represent a smoking gun of an exotic process, since it could not be explained by any 
known mechanism~\cite{dm_cap_nu_PhysRevD.34.2206, JUNGMAN1996}. 
The direction of the muons produced by the conversion of such neutrinos in the vicinity of a Cherenkov detector such as Super--Kamiokande or IceCube is strongly correlated to that of the source, providing a powerful background subtraction method~\cite{Super-Kamiokande:2004pou, Super-Kamiokande:2015xms, IceCube:2016dgk}.

Both DD and capture are substantially modified in the scenario of Inelastic Dark Matter (IDM). In this class of models a DM particle $\chi$ of mass $m_\chi$ interacts with atomic nuclei exclusively by up--scattering to a second heavier state $\chi^{\prime}$ with mass $m_{\chi^{\prime}}$ = $m_\chi$ + $\delta$. A peculiar feature of IDM is that there is a minimal WIMP incoming
speed in the target frame matching the kinematic threshold for inelastic upscatters and given by:

\begin{equation}
    v^*_{min}=\sqrt{\frac{2\delta}{\mu_{\chi N}}},
    \label{eq:vstar}
\end{equation}

\noindent with $\mu_{\chi N}$ the WIMP--nucleus reduced mass. The WIMP speed in the halo of our Galaxy is bounded by the escape velocity $u_{esc}\simeq$ 550 km/s and for both DD and capture in the Sun or the Earth the targets move in the Galactic rest frame with a relative speed $v_0\simeq$ 220 km/s corresponding to the rotational curve at the solar system position~\cite{SHM_Green:2011}.  As a consequence the WIMP-scattering process in DD vanishes if $v^*_{min}\gsim v_{max} = u_{esc} + v_0\simeq$ 800 km/s, because it becomes kinematically not accessible, the incoming WIMP speed being too slow to overcome the inelasticity threshold. Such kinematic bound implies that the largest values of $\delta$ that DD can probe are reached by the heaviest targets (xenon, iodine, tungsten) employed in existing experiments and cannot exceed $\delta\simeq$ 200 keV~\cite{iDM_DD_Nagata2014}.

A trivial way to extend the sensitivity to the $\delta$ parameter beyond DD is to look for processes where the WIMP scatters off nuclei at higher speeds.  For instance, the gravitational potential accelerates the WIMP particles before they scatter off the nuclei of a celestial body, so that their speed can be larger than $v_{max}$. Indeed, it can reach 1600 km/s in the center of the Sun, implying that the values of $\delta$ that can be probed in IDM by capture can reach $\simeq$ 600 keV~\cite{idm_sun_Catena_2018}. As a consequence, for these scenarios the solar capture bounds are stronger than those from DD.

Clearly, even stronger bounds can in principle be obtained from WIMP capture in celestial bodies with a gravitational potential stronger than that of the Sun, such as 
white dwarves (WDs)~\cite{idm_wd_2010, idm_wd_2010_Hooper, higgino_wd_2019} and neutron stars (NSs)~\cite{IDM_neutron_stars_2018, IDM_neutron_stars_2022}. Indeed, in such cases the speed of the scattering WIMP can reach $\simeq$ a few $10^4$ km/s (for WDs)~\cite{improved_WD_2021} or a few $10^5$ km/s (for NSs)~\cite{improved_neutron_stars_2020}, potentially extending the values of $\delta$ that can be probed up to several tens of MeV for WDs and hundreds of MeV for NSs.
WDs have already been used in the literature to put constraints on other DM scenarios~\cite{Cermeno:2018qgu}.

In Ref.~\cite{improved_WD_2021} the WD equation of state improved in~\cite{WD_improved_eos} was used to constrain the WIMP--nucleus cross-section in the case of elastic scattering. The aim of our paper is to extend the analysis of Ref.~\cite{improved_WD_2021} to the case of inelastic scattering, showing how this can have dramatic consequences on a specific realization of the IDM scenario, bi--doublet DM in Left-Right symmetric models (LRSM)~\cite{Garcia-Cely_LR_pheno}.

In such scenario the DM particle
has a non--vanishing hypercharge leading to a coupling to the Z boson whose size is several orders of magnitude larger than that excluded by DD experiments in the case of elastic scattering.
However it can be reconciled to DD constraints by selecting the parameter space for which the mass splitting $\delta\gsim$ 200 keV so that the nuclear scattering process is kinematically forbidden~\cite{path_to_SO10_2019}. In particular for bi--doublet DM  in the LRSM only DD constraints have been discuss so far. In Section~\ref{sec:LR} we will update its phenomenological analysis including the improved bound from WDs.

The paper is organized as follows. In Section~\ref{sec:capture} we extend the WD bounds discussed in~\cite{improved_WD_2021} to IDM. In particular in Section~\ref{sec:wd_in_gc} we review the present knowledge on the amount of DM in the globular clusters (GCs) of our Galaxy and specifically in Messier 4 (M4) showing that N--body simulations suggest that it can be large in their core. Based on this assumption, in Section~\ref{sec:capture_in_wd} we will use the data collected by the Hubble Space Telescope on low--temperature WDs in M4 to constraint IDM, showing that indeed it is possible to probe in this way values of $\delta$ as large as 30 or 40 MeV. At the end of the same Section we also provide a short discussion on how the detection of low--temperature neutron stars could further improve the bounds from WDs to values of the mass splitting $\delta$ as large as $\simeq$ 300 MeV. In Section~\ref{sec:LR} we summarize the main features of the bi--doublet LRSM model and show how the results of Section~\ref{sec:capture_in_wd} substantially reduce the range of variation of its parameters, in particular requiring a significant splitting at low energy between the couplings $g_L$ and $g_R$ of the two groups $SU(2)_L$ and $SU(2)_R$ in order to yield a thermal relic abundance in agreement to observation. In Section~\ref{sec:high_scale} we discuss the consequences of the results of Section~\ref{sec:LR} on the running of $g_L$ and $g_R$ and on their unification scale when the LRSM is embedded in a Grand Unified Theory (GUT) and provide a specific realization of such scenario, whose features are given in Appendix~\ref{app:beta_functions}.  Finally, Section~\ref{sec:conclusions} contains our Conclusions.

\section{Constraints on Inelastic Dark Matter from White Dwarves}
\label{sec:capture}

\subsection{The Dark Matter Content of the M4 Globular Cluster}
\label{sec:wd_in_gc}
The composition of massive White Dwarves is a matter of debate. In particular, recent computations of their interior suggest that their core could be composed of neon~\cite{neon_WD}. Depending on the metallicity of their surrounding medium their chemical composition can also include Carbon and Oxygen.

They are the most abundant stellar remnants, providing a promising means of probing DM interactions, complementary to terrestrial searches~\cite{dm_cap_WD_Dasgupta2019, improved_WD_2021}. In particular the core temperature of a WD is too low to ignite nuclear fusion reactions, and as a result, WDs have no internal energy source. On the other hand, compared to the Sun their interior density can be eight orders of magnitude larger~\cite{improved_WD_2021}, while their local environment can be richer of DM particles with a lower velocity dispersion, as is the case for the M4 globular cluster, where low--temperature WDs have been observed~\cite{idm_wd_2010}. All these features imply that indeed WDs in M4 can capture WIMPs more efficiently than the Sun, and that the DM annihilation process inside the WD can easily dominate its total luminosity, driving its temperature beyond the observed ones. As a consequence, as shown in Section~\ref{sec:capture_in_wd}, the observation of low--temperature WDs in M4 can yield to constraints that are stronger compared to those obtained from  WIMP capture in the Sun.

In order to set such limits it is necessary to estimate the DM density surrounding the WDs in the 
inner region of M4. Unfortunately, direct measurements of the velocity dispersion of stars within globular clusters~\cite{dm_upper_bounds_in_gc_2022} do not allow to reconstruct their gravitational potential with enough precision to obtain their DM content. As a consequence, one needs to rely on models of GC formation based on $N$--body simulations~\cite{dm_in_gc_2005,idm_wd_2010}. This clearly represents the major source of uncertainty in our analysis and other similar ones in the literature. It is fair to say that, for instance, in~\cite{dm_stars_Bertone2007}, the presence of DM in GC is referred to as a ``controversial" issue, mainly due to the effect of tidal disruption of the host galaxy. However, several analyses suggest that the cores of DM halos in globular clusters should indeed survive successive tidal interactions with the host galaxy. The analysis of the following Sections will show that if indeed M4 is rich of DM this can have dramatic consequences on some classes of particle--physics models of Dark Matter. So, following previous phenomenological analyses~\cite{idm_wd_2010, dm_cap_WD_Dasgupta2019, improved_WD_2021} we will assume this possibility.  In the present paper we will use the results of~\cite{idm_wd_2010}. We summarize such results below, addressing the reader to the original literature for further details.

An estimate of the DM content in the M4 GC is obtained in~\cite{idm_wd_2010} by modeling the GC formation starting from an original DM subhalo of mass $M_{\rm DM} \simeq 10^7 M_\odot$ that falls into our host galactic halo~\cite{gc_peebles_1984}. Once the GC falls into the host halo, mass loss occurs continually through tidal stripping, and the orbit of the subhalo decays down toward the center. Such mass loss can be significant, resulting in only a few percent of the DM of the original subhalo surviving inside the final GC, with a baryon-dominated core with a small mass-to-light ratio that actually resembles those of observed globular clusters. The presence of baryons in the GC must also be taken into account: on the one hand, in the stellar core it can  gravitationally enhance the DM density, while on the other it can reduce it, as a consequence of the heating due to the interaction of DM particles with stars (albeit this latter effect is estimated to be negligible at the radii where WDs are observed within M4). Moreover, when the gas in the original halo which eventually forms the globular cluster loses energy and falls into the core conservation of angular momentum is expected to lead to a contraction of the DM core: also such effect turns out to be negligible at the position where the WDs are observed in M4. Finally, the effect of tidal stripping is taken into account by truncating the density distributions of stars and DM at the tidal radius of the GC. The baryon and DM density profiles within M4 ensuing from such procedure are shown in Fig.~1 of \cite{idm_wd_2010}, which implies that DM makes up less then $\simeq$ 40\% of the total mass of the cluster, in agreement to observations that indicate a lack of DM in globular clusters. In this sense the authors of~\cite{idm_wd_2010} consider such estimation to be conservative. In particular such value corresponds to less than 1\% of the original $10^7 M_{\odot}$ halo. Moreover in Ref.~\cite{idm_wd_2010} the value of the DM density at the largest radius where WDs within M4 are observed ($\simeq$ 2.3 pc, well inside the tidal radius) is estimated to be $\rho_{\rm DM}\simeq$21 $M_\odot$ pc$^{-3}$ = 798 GeV cm$^{-3}$. Furthermore, using the baryon and DM density profiles estimated with such procedure a WIMP velocity dispersion of $\simeq 8$ km/s can be obtained by assuming hydrostatic equilibrium and spherical symmetry.

The quantitative estimations of Ref.~\cite{idm_wd_2010} and summarized in this Section will be used in Section~\ref{sec:capture_in_wd} to evaluate the effect of DM capture on the WDs within M4.

\subsection{WIMP Capture in White Dwarves}
\label{sec:capture_in_wd}

Signals generated by pair annihilations of WIMPs captured and accumulated inside the Earth and the Sun have been discussed for a 
long time~\cite{capture_Press_Spergel, capture_Gould_1987ApJ, capture_Gould_1988ApJ, gould_capture_1991}. 

In the case of the Earth and the Sun only neutrinos escape from the celestial body, that can be searched for in the up-going muons flux generated by their conversion in the rock close to a Cherenkov detector. Such signals are sensitive to the yield of the neutrinos produced above the detector's threshold by the particles produced in the WIMP annihilation, typically fermions or gauge bosons, that are boosted to an energy equal to the WIMP mass before decaying. An exception to this mechanism is provided by light quarks, which are stopped inside the medium before decaying, producing below--threshold neutrinos that are not detectable. 

In compact stars such as white dwarves or neutron stars the capture process is analogous. However, in this case the observable effect is an increase of the total luminosity or of the temperature of the star, which depends only on the total electromagnetic energy $E_\gamma$ injected by the annihilation process. 
In particular, for the high energies corresponding to the values of the WIMP mass discussed in Section~\ref{sec:LR}, also the the neutrinos produced in the annihilation process are converted into electromagnetic radiation, so that the signal can be practically considered as bolometric and $E_\gamma$ = 2 $m_\chi$.

While crossing a compact star DM particles may get trapped in its gravitational potential, continue to lose energy by scattering against its constituents and get ultimately captured. A DM particle $\chi$ (of mass $m_{\chi}$) can undergo an inelastic transition to a second heavier state (say, $\chi^{\prime}$) by scattering 
against a nucleus (of mass $m_N$) present inside the star if the total kinetic energy in the DM-nucleus center of mass frame is larger than the mass difference $\delta$ between the DM 
particle and the heavier state~\cite{idm_sun_Nussinov_2009, idm_sun_Menon_2009, idm_sun_Shu_2010, idm_sun_Catena_2018}, i.e.,
\begin{equation}
\frac{1}{2} \mu_{\chi N} w^2 > \delta,
\label{eq:KE_condition}
\end{equation}
\noindent which is equivalent to Eq.~(\ref{eq:vstar}). The quantity $w$, which is the speed of the DM particle at a radial distance $r$ from the centre of the star, is given by:
\begin{equation}
w(r) = \sqrt{u^2 + v_{esc}(r)^2},
\end{equation}
where $v_{esc}(r)$ is the local escape velocity from the star at radius $r$ and $u$ is the asymptotic 
speed of the DM particles far away from the star. 

Considering the WD stars in a nearby globular cluster such as M4, a common choice for the distribution of the speed $u$ in the cluster is a boosted Maxwell Boltzmann (MB) one, which in the 
reference frame of a WD can be expressed as:
\begin{equation}
f(u) = \frac{u}{v_d v_*} \sqrt{\frac{3}{2 \pi}} \left({\rm exp} \left[-\frac{3}{2 v^2_d}(u-v_*)^2\right] - 
{\rm exp} \left[-\frac{3}{2 v^2_d}(u+v_*)^2\right]\right) ,
\label{eq:MB_velocity}
\end{equation}
where $v_*$ and $v_d$ are the velocity of the WD and the velocity dispersion of DM 
inside the globular cluster, respectively. 
Following~\cite{idm_wd_2010, improved_WD_2021}, we have assumed conservative values for $v_*$ and $v_d$, which are $v_* = 20$ km/s and $v_d = 8$ km/s for the WDs in M4.

Given the density profiles of nuclear elements in a WD, one can estimate the corresponding escape velocity profile $v_{esc}(r)$. In particular, the authors of Ref.~\cite{improved_WD_2021} have used an improved equation of state for WDs and calculated the density profiles of thermal electrons and nuclear ions for WDs with different configurations that are made of elements like carbon, oxygen, etc. In particular assuming that the vast majority of heavier WDs are made of carbon, oxygen or neon, carbon provides the lowest rate for a spin--independent (SI) cross section that scales with the atomic mass number squared. Indeed, this is the most common type of interaction expected for WIMPs with nuclei, and it arises in the specific scenario analyzed in Section~\ref{sec:LR}, so in the following we will assume that the WDs are entirely made of carbon to get conservative bounds. 
Among the various benchmarks for carbon WD tabulated in~\cite{improved_WD_2021} we take the heaviest WDs 
with masses $M_* = 1.38$ \(M_\odot\) and 1.25 \(M_\odot\) and extract the associated radial profiles for the nuclear density and the escape velocity. The corresponding radii of these two benchmark WDs are $R_* \simeq 1.25 \times 10^3$ km and 
$R_* \simeq 3.29 \times 10^3$ km, respectively.

For the heaviest WD considered here, the escape velocity $v_{esc}$ in the interior of the star 
can reach a very high value $\sim 0.1$ $c$~\cite{improved_WD_2021} which in turn extends the kinematic upper limit (shown in Eq.~(\ref{eq:KE_condition})) on $\delta$ up to several tens of MeV. Due to this reason, using heavier WDs it is possible to probe a high value of $\delta$, as will be described 
in detail in this sub-section.

In the optically thin limit, DM particles crossing a WD can become 
gravitationally bound to the star after a single scattering. In this case, the rate of DM capture through inelastic scattering 
is given by~\cite{idm_sun_Catena_2018, idm_sun_Shu_2010, idm_sun_Menon_2009, idm_sun_Nussinov_2009}:
\begin{equation}
C_{\rm opt-thin} = \frac{\rho_{\chi}}{m_{\chi}} \int_0^{R_*} dr \hspace{0.5mm} 4 \pi r^2 
\int_0^{\infty} du \hspace{0.5mm} \frac{f(u)}{u} \hspace{0.5mm} w \hspace{0.5mm} \Omega (w,r) 
\hspace{1mm} \Theta \left(\frac{1}{2} \mu_{\chi N} w^2 - \delta \right) .
\label{eq:C_opt_thin}
\end{equation}
Here $\rho_{\chi}$ is the density of DM at the location of the WD inside the globular cluster. 
As discussed in Section~\ref{sec:wd_in_gc} in the local environment of the WDs inside M4 we take $\rho_{\chi}$ = 798 GeV cm$^{-3}$.
Note that the $\Theta$ function in Eq.~(\ref{eq:C_opt_thin}) ensures that the capture rate goes to zero 
if the condition given in Eq.~(\ref{eq:KE_condition}) is violated.
The interaction rate $\Omega (w,r)$ can be obtained using the following expression:
\begin{equation}
\Omega (w,r) = \eta_{N}(r) \hspace{0.5mm} w \hspace{0.5mm} \Theta(E_{\rm max}-E_{\rm cap}) 
\int_{E_{\rm min}}^{E_{\rm max}} dE \hspace{0.5mm} \frac{d\sigma[\chi+N \rightarrow \chi^{\prime}+N]}{dE} \hspace{1mm} \Theta(E-E_{\rm cap}) ,
\label{eq:interaction_rate}
\end{equation} 
where $\eta_{N}(r)$ is the number density distribution 
(extracted from~\cite{improved_WD_2021} for the two benchmark WDs considered) 
of the target nuclear ion (in our case $^{12}C$) in the WD, 
and $\frac{d\sigma[\chi+N \rightarrow \chi^{\prime}+N]}{dE}$ is the differential cross-section 
(as a function of the recoil energy $E$) for the DM-nucleus inelastic scattering 
$\chi+N \rightarrow \chi^{\prime}+N$. 
 The minimum and the maximum recoil 
energies (i.e., $E_{\rm min}$ and $E_{\rm max}$) corresponding to this scattering process and 
the minimum energy transfer $E_{\rm cap}$ required for 
the capture (i.e., to scatter a DM particle from $w$ to a velocity less than $v_{esc}(r)$) are: 
\begin{eqnarray}
E_{\rm min, \rm max}&=& \frac{1}{2} m_{\chi} w^2 \left[1 - \frac{\mu^2_{\chi N}}{m^2_N} \left(1 \pm \frac{m_N}{m_{\chi}} 
\sqrt{1 - \frac{\delta}{\mu_{\chi N} w^2 / 2}}\right)^2 \right] - \delta ,
\nonumber\\
E_{\rm cap}&=& \frac{1}{2} m_{\chi} u^2 - \delta .
\label{eq:E_max_min_cap}
\end{eqnarray}
In Eq.~(\ref{eq:interaction_rate}) the first Heaviside step function $\Theta$ imposes the condition 
$E_{\rm max} > E_{\rm cap}$ which is required for the capture to occur, while the second one sets the lower limit of the energy integration 
to ${\rm max}[E_{\rm min}, E_{\rm cap}]$.

Notice that, after the first up-scattering of the DM particle $\chi$, 
there are two possible situations that may arise depending on the lifetime of the heavier state $\chi^{\prime}$. If the lifetime of $\chi^{\prime}$ is very short, it quickly decays back to $\chi$ along with other particle(s) which are much lighter and carry away most of the energy produced in the decay, so that no significant kinetic energy is imparted to $\chi$. As a consequence Eqs.~(\ref{eq:C_opt_thin}, \ref{eq:interaction_rate}) can be used to calculate the capture rate because a single scattering with $E>E_{\rm cap}$ is sufficient to keep the $\chi$ particle gravitationally bound during the lifetime of the star eventually driving it to the core of the celestial body after subsequent interactions with its nuclear targets. We will see that this is the situation which occurs in the types of DM models discussed in the next Section. On the other hand, if the lifetime of $\chi^{\prime}$ is large enough it eventually down-scatters off a WD nuclear target back to a $\chi$ through an exothermic process that can in principle eject the outgoing DM particle to a non--bound orbit. In this case Eqs.~(\ref{eq:C_opt_thin}, \ref{eq:interaction_rate}) cannot be used to describe the capture process. However, if $\chi$ is much heavier than the target nucleus, the latter carries away most of the energy, leading to a situation where 
the DM particle does not have sufficient kinetic energy to escape from the star's gravitational potential and thus also in this case it is eventually captured~\cite{idm_sun_Menon_2009}. Actually, in the present work we will be interested in the capture of TeV scale DM particles, so they would be captured in the WD star after the first inelastic scattering with $E>E_{\rm cap}$ even for $\chi^{\prime}$ lifetimes longer than those discussed in Section~\ref{sec:LR_WD_bounds}.

The optically thin approximation holds when the 
scattering cross-section (i.e., the integration of $\frac{d\sigma}{dE}$ over $E$ in Eq.~(\ref{eq:interaction_rate})) is small enough so that the effect of multiple scatterings can be neglected. If the cross-section becomes very large, each DM particle that traverses the WD
is captured and the capture rate saturates to its maximum value (known as the geometric limit) 
which is independent of the DM-nucleus interaction~\cite{improved_WD_2021, dm_cap_sun_Garani2017, idm_wd_2010_Hooper, dm_stars_Bertone2007, dm_cap_sun_Bottino2002}:
\begin{equation}
C_{\rm geom} = \pi R^2_* \left( \frac{\rho_{\chi}}{m_{\chi}} \right) \int_0^{\infty} 
du \hspace{0.5mm} \frac{f(u)}{u} \hspace{0.5mm} w^2(R_*).
\label{eq:C_geom}
\end{equation}

Finally, considering both regimes (i.e., optically thin and geometric regimes), 
the capture rate of DM in a WD star through inelastic scattering 
can be estimated as~\cite{higgino_wd_2019, dm_cap_sun_Garani2017, dm_cap_sun_Bernal2012, dm_stars_Bertone2007}: 
\begin{equation}
C_{*} = {\rm min} [C_{\rm opt-thin}, C_{\rm geom}] .
\label{eq:C_approx}
\end{equation}

After being captured, the DM particles continue to scatter with the constituents of the WD and eventually thermalise and settle down to the WD interior where they can annihilate in pairs 
to produce various SM particles. The thermalisation time-scale is expected to be several orders of magnitude smaller than the 
typical age of the old and cold WDs (the types of which have been considered here) in the M4 globular cluster~\cite{improved_WD_2021}. 
The evolution of the number of DM particles ($N_{\chi}$) in the WD interior can be 
expressed as~\cite{improved_WD_2021}, 
\begin{equation}
\frac{dN_{\chi}}{dt} = C_* - A N^2_{\chi} ,
\label{eq:dNchi_dt}
\end{equation}
where $A$ is the annihilation coefficient (which is proportional to 
the averaged annihilation cross-section $\langle \sigma v \rangle$) and is related to the annihilation rate as, 
\begin{equation}
\Gamma_{\rm ann} = \frac{1}{2} A N^2_{\chi} .
\label{eq:ann_rate}
\end{equation}

The primary SM particles injected by DM annihilation in a WD interior lead to further cascades and give rise to secondary particles which quickly thermalise in the medium of the star and thus end up in heating the star. Assuming that the total energy emitted in the DM annihilation process contributes to heat up the star, the resulting luminosity can be estimated as: 
\begin{equation}
L_{\chi} = 2 m_{\chi} \Gamma_{\rm ann} .
\label{eq:L_chi0}
\end{equation}

Note that, in reality, some percentage of the energy produced in DM annihilation inside the WD may be carried away by neutrinos which escape the star without heating it up. 
However, we have checked that, for the WDs considered here, 
neutrinos produced in the star core with energies above a GeV cannot come out from the star 
as their mean free path, estimated using the number density of the WD constituents 
(obtained in~\cite{improved_WD_2021} considering an improved analysis) 
and the interaction cross-section of neutrinos with those constituents~\cite{nu_cross-section_2012}, is much smaller than the typical size of the WD star. The energetic neutrinos that cannot escape the star eventually thermalise in the star and are converted into electromagnetic radiation.
In case of the annihilation of a TeV scale DM (which is our present interest), 
the fraction of energy carried away by neutrinos with energies below 1 GeV (that are produced in the cascade of various primary annihilation states such as $W^+W^-, ZZ, b\bar{b}$) 
is less than 1$-$2\% (estimated using~\cite{Cirelli_PPPC}).
Thus, we can assume that almost all of the energy produced in the annihilation contributes to the luminosity of the star and use Eq.~(\ref{eq:L_chi0}) to calculate it. Notice that this implies that the ensuing bounds do not depend on assumptions on the primary SM states the DM particles annihilate to. This is at variance with capture in the Sun or the Earth that may lead to a neutrino flux below the experimental energy threshold when the DM particles annihilate only to light states that are stopped before decaying.

The capture and annihilation processes inside the WD star equilibrate (so that $\frac{dN_{\chi}}{dt}$ becomes zero) over a time-scale $\tau_{\rm eq} = (C_* A)^{-1/2}$. 
A conservative estimate (following~\cite{higgino_wd_2019}) with the assumptions that 
$\langle \sigma v \rangle = \langle \sigma v \rangle_{\rm thermal}$ and $C_* = C_{\rm geom}$, 
the equilibrium time-scale $\tau_{\rm eq}$ (for DM particles of mass $m_{\chi} = 1$ TeV) 
turns out to be of the order of $\sim 10^{-6}$ Gyr for our benchmark WDs. 
This time-scale is clearly several orders of magnitude lower than the typical ages of the old and faint WDs in M4, which are expected to be at least a few Gyrs~\cite{improved_WD_2021}. 
Hence, one can in general assume equilibrium between the capture and annihilation processes inside these WDs in M4, and specifically in the model discussed in Section~\ref{sec:LR}. 
In that case the DM annihilation rate is $\Gamma_{\rm ann} = C_* / 2$ and the corresponding luminosity (given by Eq.~(\ref{eq:L_chi0})) becomes~\cite{idm_wd_2010_Hooper, dm_cap_WD_Dasgupta2019, improved_WD_2021}: 
\begin{equation}
L_{\chi} = m_{\chi} C_* .
\label{eq:L_chi}
\end{equation}

\begin{figure*}[ht!]
\centering
\includegraphics[width=7.49cm,height=7.4cm]{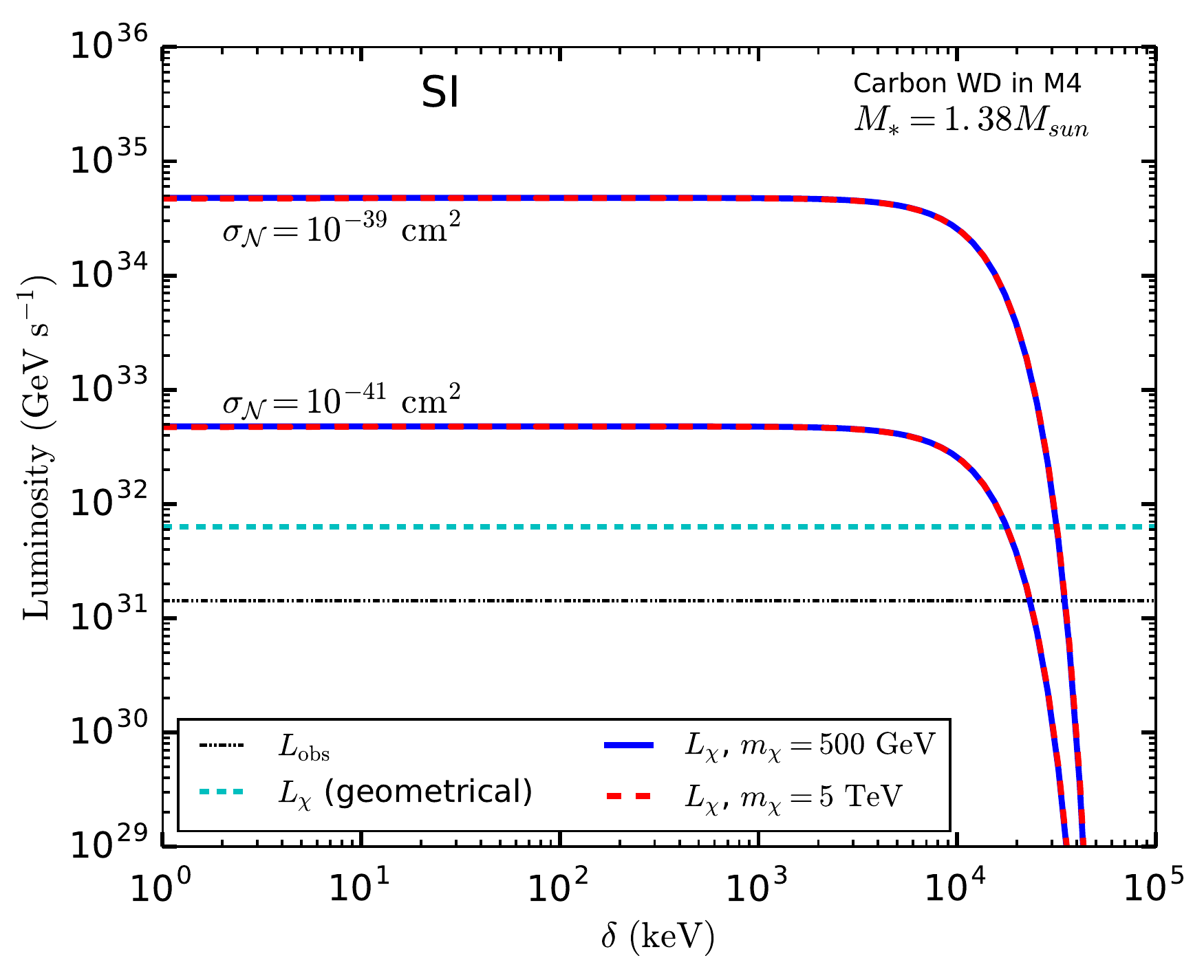}
\includegraphics[width=7.49cm,height=7.4cm]{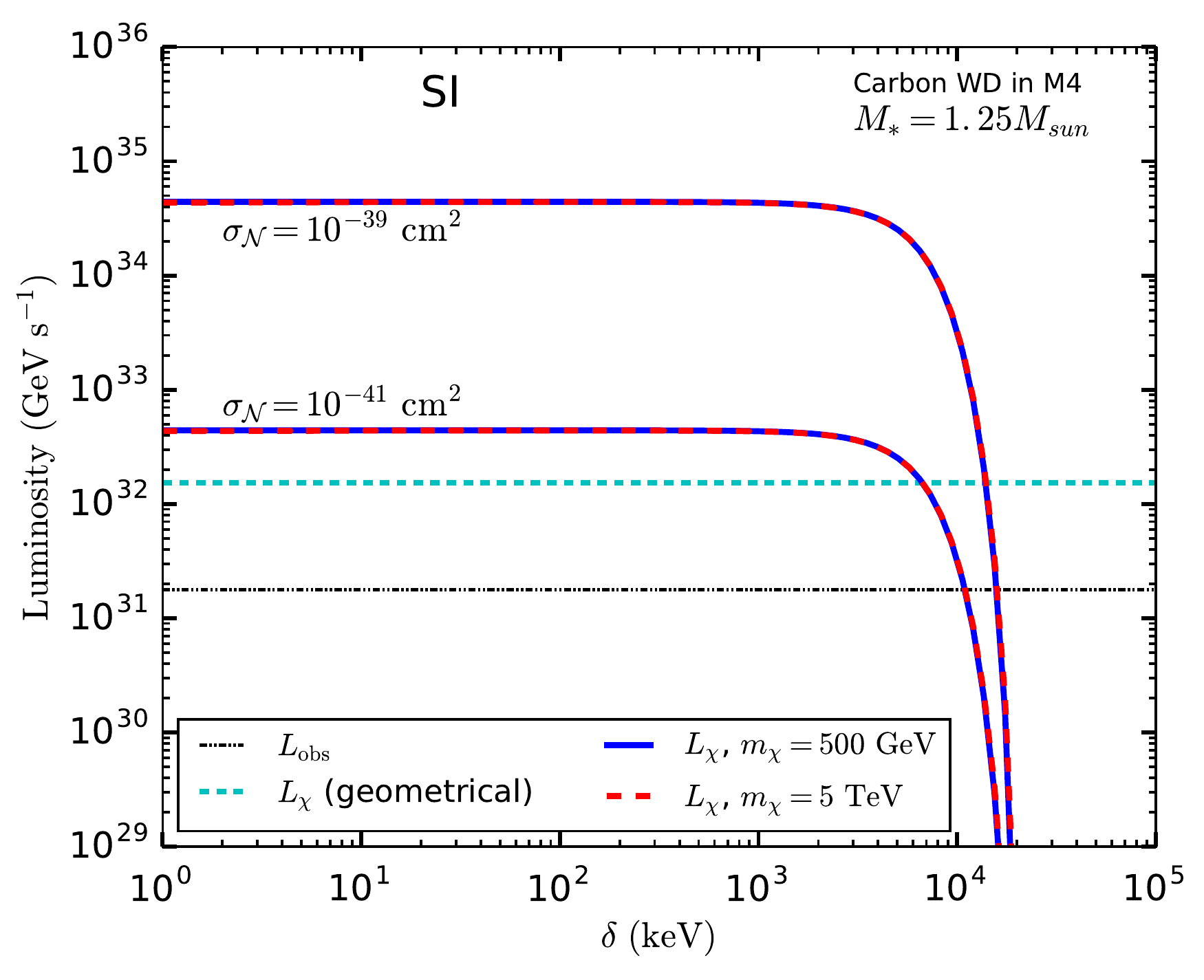}
\caption{DM induced luminosities in both optically thin (blue and red lines) and geometric (cyan lines) limits as functions of $\delta$, considering two benchmark carbon WDs in M4 
with masses $M_* = 1.38$ \(M_\odot\) (left panel) and $M_* = 1.25$ \(M_\odot\) (right panel). 
The blue and the red lines are corresponding to the capture of 500 GeV and 5 TeV DM particles, respectively, assuming a spin-independent DM-nucleus interaction. 
In both panels the black horizontal lines indicate the luminosities (taken from~\cite{improved_WD_2021}) that are observed for the considered WDs, while
the cyan horizontal lines show the luminosities of the two benchmark WDs when DM capture is calculated in the geometric limit.}
\label{fig:cap_Lumi_SI}
\end{figure*}

In Fig.~\ref{fig:cap_Lumi_SI}, we show, as a function of the mass-splitting $\delta$, the luminosities that are induced by the annihilation of inelastically captured DM particles in carbon WDs (located in M4) with masses and radii
$M_* = 1.38$ \(M_\odot\), $R_* \simeq 1.25 \times 10^3$ km (left panel) and 
$M_* = 1.25$ \(M_\odot\), $R_* \simeq 3.29 \times 10^3$ km (right panel).   
The blue and the red lines in each panel represent the luminosities 
corresponding to the capture of DM with masses $m_{\chi} = 500$ GeV and 5 TeV, respectively, assuming a standard SI cross-section with an exponential (Helm) nuclear form factor~\cite{dm_cap_sun_PhysRevD.86.073002, idm_sun_Shu_2010} for the WIMP scattering events off carbon ($^{12}C$) in the optically thin limit (Eq.~(\ref{eq:C_opt_thin})). In particular, two illustrative values of the DM-nucleon cross-section $\sigma_{\mathcal{N}}$ 
have been used. The cyan horizontal lines in both panels of Fig.~\ref{fig:cap_Lumi_SI} 
show the luminosities obtained for the two benchmark WDs 
considering the capture of DM in the geometric limit (Eq.~(\ref{eq:C_geom})).
Note that, since the capture rate falls as the inverse of $m_{\chi}$ in the geometric limit 
as well as in the optically thin limit (for $m_{\chi}\lesssim {\cal O}(10^5)$ GeV~\cite{improved_WD_2021} and much larger than the target nucleus mass), the estimated luminosities in both regimes are independent of the DM mass.

The additional black lines shown in Fig.~\ref{fig:cap_Lumi_SI} represent the luminosities for the two considered WDs in M4 observed by the Hubble Space Telescope~\cite{hubble_WD_2004, hubble_WD_2009} and taken from~\cite{improved_WD_2021}.
These two WDs are among the faintest WDs observed in M4. Therefore, if the DM predicted contribution to the total luminosity of these WDs is less than or at most equal to the observed ones, then there will be no contradiction between the DM prediction and the observations.

There are a few important points to notice in Fig.~\ref{fig:cap_Lumi_SI}. 
The first one is that the DM luminosities calculated for the selected WDs in the optically thin limit are largely independent of $\delta$ up to a certain value of this parameter, beyond which 
the capture rates (hence the luminosities) start falling rapidly and ultimately
go to zero when $\delta$ exceeds several tens of MeV. 
The second point is that the maximally achievable DM luminosity (i.e., the luminosity obtained in the geometric limit) for each WD is above the observed limit. 
Based on these two points, a comparison between the WD data and the DM luminosity (obtained using Eqs.~(\ref{eq:C_approx}) and (\ref{eq:L_chi})) enables one to constrain the DM interaction strength (for example, the DM-nucleon cross-section $\sigma_{\mathcal{N}}$ 
in the case of a SI interaction) 
up to a maximum value of $\delta$ (a few tens of MeV for the WDs considered here), 
above which the DM luminosity goes below the observation. 
The maximum value of $\delta$ depends on the WD parameters 
and is larger for heavier WDs, since heavier and compact WDs have stronger gravitational potential which increases the kinematic upper limit on $\delta$ 
(see Eq.~(\ref{eq:KE_condition})). Notice that the mass of the heaviest of our benchmark WDs
almost saturates the Chandrasekhar limit 
$M_*\lsim$ 1.4 \(M_\odot\)~\cite{Chandrasekhar_limit, Chandrasekhar_limit_2007}, 
and so in practice it kinematically saturates the 
best achievable bound on $\delta$ using WDs.

\begin{figure*}[ht!]
\centering
\includegraphics[width=7.49cm,height=7.4cm]{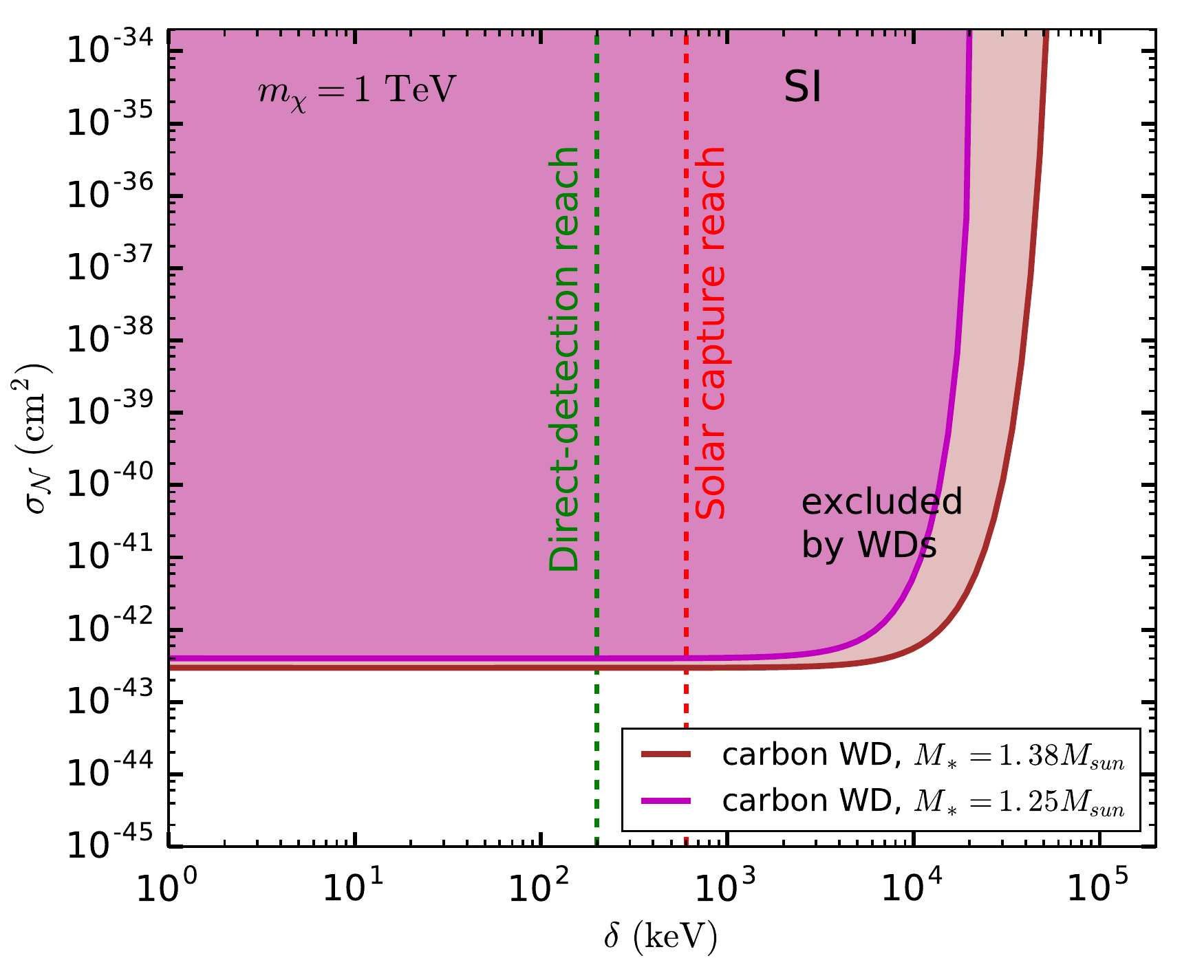}
\caption{Constraints in the $\sigma_{\mathcal{N}}$ vs. $\delta$ plane (for a SI interaction) obtained using the observed luminosities for the two selected WDs in M4. The dashed green and red vertical lines  show the maximum values of $\delta$ that can be probed in present direct-detection experiments 
and through solar neutrino searches, respectively.}
\label{fig:sigma_del_SI}
\end{figure*}

In order to illustrate the above discussion, we show in Fig.~\ref{fig:sigma_del_SI} the regions of the $\sigma_{\mathcal{N}}$ vs. $\delta$ plane (for a SI interaction) that are excluded using the luminosity data of two selected faint WDs located in M4, assuming that they are made of carbon. The DM mass is taken to be $m_{\chi} = 1$ TeV, although, as mentioned earlier, for a sufficiently heavy DM (much heavier than the target nucleus) these constraints do not depend on $m_{\chi}$. 
In the same figure the green and the red dashed lines indicate the maximum values of $\delta$ that can be probed in direct-detection experiments~\cite{iDM_DD_Nagata2014} and in solar neutrino searches~\cite{idm_sun_Catena_2018}, respectively. 
Fig.~\ref{fig:sigma_del_SI} shows that WDs can improve the pre--existing bounds on $\delta$ by more than one order of magnitude ruling out $\delta$ up to a few tens of MeV for $\sigma_{\mathcal{N}}\gsim$ 3$\times$ 10$^{-43}$ cm$^2$.

The constraints that we obtain here using the observations of faint and heavy WDs in M4 assume a spin-independent interaction between DM and the target nucleus. WDs are expected to be mostly composed of spinless targets ($^{12}C$, $^{16}O$ or $^{20}Ne$)\footnote{The abundance of targets with spin such as $^{13}C$ and $^{19}O$ in WDs is however subject to large uncertainties and strongly depends on the WD progenitor~\cite{C13_in_WD_2011}.}, so that in order to derive constraints on spin-dependent (SD) type interactions one needs to consider DM capture in other celestial objects such as the Sun. Notice that, as already pointed out ,the reach on $\delta$ in the case of the Sun is considerably smaller (up to $\simeq$ 300 keV for SD)~\cite{idm_sun_Shu_2010, idm_sun_Catena_2018}.

Apart from WDs, one can also consider capture of DM particles in more compact objects 
such as neutron stars (NSs)~\cite{IDM_neutron_stars_Kouvaris2007, IDM_neutron_stars_2018, improved_neutron_stars_2020, IDM_neutron_stars_2022} 
where the gravitational potential is so strong that the infalling DM particles are accelerated to quasi-relativistic speeds. When such a high speed DM particle collides with the NS constituents it
can up-scatter to the heavier state for a value of the inelastic splitting $\delta$ 
that can be as large as $\sim 300$ MeV~\cite{IDM_neutron_stars_2018, IDM_neutron_stars_2022}. 
The capture process of DM particles and their subsequent annihilation at the star core can 
heat a NS maximally up to a temperature $T^{\infty}$ (measured at a large distance
from the NS) which is given by~\cite{IDM_neutron_stars_2018, dm_cap_NS_Busoni2021, dm_capture_NS_Chatterjee2022}: 
\begin{equation}
T^{\infty} \simeq 2300 \hspace{0.6mm} {\rm K} 
\left(\frac{\rho_{\chi}}{0.4 \hspace{0.6mm} {\rm GeV cm^{-3}}} \right)^{1/4} 
F \left(\frac{v_*}{230 \hspace{0.6mm} {\rm km \hspace{0.5mm} s^{-1}}} \right) , 
\label{eq:T_NS}
\end{equation}
where $\rho_{\chi}$ is the DM density at the location of the NS, $v_*$ is the velocity of the NS in the halo and the function $F$ is defined as,
\begin{equation}
F \left( x \right) = 
\left[ \frac{{\rm Erf}(x)}{x \hspace{0.6mm} {\rm Erf}(1)} \right]^{1/4} .
\label{eq:err_fn}
\end{equation}
Eq.~(\ref{eq:T_NS}) corresponds to a temperature $\simeq$ a 
few$\times 10^3$ K for nearby NSs. 
The observation of old NSs with such a low temperature is beyond the reach of existing telescopes but could be achievable by forthcoming infrared telescopes such as the James Webb Space Telescope (JWST), the Thirty Meter Telescope (TMT) and the European Extremely Large Telescope (E-ELT)~\cite{NS_infrared_Baryakhtar2017, dm_capture_NS_Chatterjee2022, NS_infrared_PhysRevD.97.043006}.

\section{The case of Bi--doublet DM in Left--Right Symmetric Models}
\label{sec:LR}
The phenomenological discussion of Section~\ref{sec:capture} shows how WDs can provide bounds on the IDM scenario that are significantly stronger compared
to those from DD. In this Section we wish to illustrate the impact that such enhanced bounds can have on a specific IDM model, the fermion bi-doublet extension of the Left--Right symmetric model (LRSM). 

The main motivation of the LRSM scenario is to explain one of the most puzzling features of the Standard Model, i.e. the observed maximal parity violation in its weak sector that forces to arrange the left--handed chiral components of fermions in $SU(2)$ electroweak doublets and their right--handed counterparts in singlets. In the LRSM scenario the SM gauge group is enlarged in order to contain an $SU(2)_L$ and an $SU(2)_R$, so that the doublets of $SU(2)_L$ are singlets of $SU(2)_R$ and the doublets of $SU(2)_R$ are singlets of $SU(2)_L$.        
Interestingly, such scenarios can be easily extended by a multiplet that contains a Dark Matter candidate that is automatically stable~\cite{Cirelli:2005uq, Heeck_2015_LRSM_DM_stability} and whose mass is the only required additional free parameter. A detailed description of the LRSM scenario can be found in \cite{Duka:1999uc}, and a discussion of the phenomenology of its DM extensions in~\cite{Garcia-Cely_LR_pheno}. In Section~\ref{sec:model_description} we will summarize the specific scenario where the LRSM model is extended with a fermion bi-doublet that contains an IDM candidate, and in Section~\ref{sec:LR_WD_bounds} we apply the results of  Section~\ref{sec:capture_in_wd} to constrain its parameter space. Moreover, in Section~\ref{sec:high_scale}  we discuss how the low-energy LRSM parameter space compatible with observation can be obtained when the LRSM model is embedded in a Grand Unified Theory at high energy.

\subsection{Model Description}
\label{sec:model_description}

The gauge symmetry of LRSM  is $SU(3)\times SU(2)_{L} \times {SU}(2)_R \times {U(1)}_{B-L}$, with $B$ and $L$ the baryon and lepton number. The representations under the different gauge groups of the fermions and scalars contained in the model are listed in Tab.~\ref{tb:LRSM_MC} of Appendix~\ref{app:beta_functions} for both its minimal realization and for various extension that will be discussed in Section~\ref{sec:high_scale}. 
In its minimal realization the LRSM preserves D parity, i.e. it is invariant by the interchange of any multiplet of $SU(2)_L$ into the corresponding  $SU(2)_R$ multiplet. In particular, in one of the most popular variants of minimal LRSM the scalar sector
contains an $SU(2)_L$ triplet $T_L$, an $SU(2)_R$ triplet
$T_R$ and a scalar bi-doublet $\Phi$. The LR-symmetry of the model is broken spontaneously down to the SM in two steps. First, the $SU(2)_R\times U(1)_{B-L}$ symmetry is broken spontaneously to $U(1)_{Y}$ at a high scale $M_R$ by the vacuum expectation value (VEV) $v_R$ of $T_R$, and the masses of the $SU(2)_R$ gauge bosons $W_R$ and $Z_R$ are generated. This breaks the parity invariance of the model and implies the relation $Y = T^3_R + \dfrac{1}{2}(B-L)$ between the generator of  $U(1)_Y$ and those of $U(1)_{B-L}$ and $SU(2)_R$.
In particular $Z_{R}$ is a linear combination of $W^3_R$ (the gauge boson corresponding to the generator
$T^3_R$) and $B$ (the gauge boson for $U(1)_{B-L}$), while the orthogonal
combination $B_Y$ (the gauge boson corresponding to the remnant $U(1)_Y$ symmetry) remains massless at this stage. Notice that the breaking of
$SU(2)_R\times U(1)_{B-L}$ also generates large Majorana masses for right--handed neutrinos.
The second stage of symmetry breaking happens at the electroweak scale, when the scalar
bi-doublet $\Phi$ gets VEVs $v_1/\sqrt{2}$ and $v_2/\sqrt{2}$, and the intermediate $SU(2)_{L}\times U(1)_Y$ breaks down to
the $U(1)_{em}$ symmetry, leading to the usual relation $Q = T^3_L + Y$, so that $Q= T^3_L + T^3_R +  \dfrac{1}{2}(B-L)$. This leads to the mass generation
of all the known SM particles. Moreover, since $\Phi$ is a bi-doublet, its VEVs generate mixings among left--handed and 
right--handed gauge bosons. This also relates the gauge couplings in the following manner\footnote{When models are embedded into a Grand Unified Theory, the normalization of $g_R$ and $g_{B-L}$ is that of Eqs.~(\ref{eq:matchingtoSM}-\ref{eq:matchingtoSMBLMR}).}
\begin{eqnarray}
\dfrac{1}{g_{Y}^2} &=& \dfrac{1}{g^2_R} + \dfrac{1}{g^2_{B-L}} \,,\\
\dfrac{1}{e^2} &=& \dfrac{1}{g^2_L} + \dfrac{1}{g^2_Y} \,,
\end{eqnarray}
where $g_{\alpha}$ is the gauge coupling of
$U(1)_{\alpha}$ ($\alpha=B-L,\,Y$). The $SU(2)_{L(R)}$ gauge
coupling is denoted by $g_{L(R)}$ while $e$ is the
gauge coupling of the residual $U(1)_{em}$. 
The three mixing angles among the neutral gauge
bosons i.e. $\theta_R$ (mixing between $W^{3}_R$ and $B$), $\theta_W$
(mixing between $W^{3}_L$ and $B_Y$, the Weinberg angle) and $\phi$ 
(mixing between $Z_{R}$ and $Z_{L}$) are given by~\cite{Garcia-Cely_LR_pheno}:
\begin{equation}
\theta_W = \tan^{-1}\left(\dfrac{g_Y}{g_L}\right),\,\, 
\theta_R = \sin^{-1}\left(\dfrac{g_L}{g_R}\,\tan\theta_W\right),\,\, \phi \simeq \dfrac{1}{2}\tan^{-1}\left(-2\,\cos\theta_W\cos\theta_R
\dfrac{g_R}{g_L}\dfrac{M^2_{Z_L}}{M^2_{Z_R}}\right), 
\label{theta_phi}
\end{equation}
where: 
\begin{eqnarray}
M^2_{Z_L}\simeq M^2_{Z_1} = \dfrac{g^2_L}{4\,\cos^2\theta_W}\,v^2=M^2_Z,\,\,\,
M^2_{Z_R}\simeq M^2_{Z_2} = \dfrac{g^2_R}{\cos^2\theta_R} v^2_R +
\dfrac{g^2_R}{4} \cos^2\theta_R\,v^2.
\label{eq:mixings}
\end{eqnarray}
\noindent In Eq.~(\ref{eq:mixings}) the mixing $\phi$ is neglected since $M_{Z_R}\gg M_{Z_L}$,
$Z$ is the Standard Model massive neutral gauge boson and $v=\sqrt{v^2_1+v^2_2} =246$ GeV. 
The expression of $\phi$ in Eq.~(\ref{theta_phi}) is valid in the phenomenological limit where
$v_R>>v_1,\,v_2$ and the VEV of $T_L$ ($v_L$) $\rightarrow0$. Similarly,  when $v_1$ and $v_2$ are real the mixing angle between $W^{\pm}_{R}$ and $W^{\pm}_{L}$ is given, in the phenomenological limit, by~\cite{Garcia-Cely_LR_pheno}:
\begin{eqnarray}
\xi &\simeq& \dfrac{1}{2}\tan^{-1}\left(-4\,
\dfrac{g_R}{g_L}\dfrac{M^2_{W_L}}{M^2_{W_R}}\dfrac{v_1v_2}{v^2}\right)\,, 
\label{eq:tan_xi}
\end{eqnarray}
with:
\begin{eqnarray}
M^2_{W_L}\simeq M^2_{W_1} = \dfrac{g^2_L}{4}\,v^2=M^2_W,\,\,\,\,\,\,\,\,\,\,
M^2_{W_R}\simeq M^2_{W_2} = \dfrac{1}{4} g^2_R\left(v^2+2v^2_R\right)\,,
\end{eqnarray}
where $M_W$ is the mass of the electroweak charged gauge boson of the Standard Model.

In the DM scenario that we wish to discuss the LRSM is minimally extended by adding a self--conjugate fermionic bi-doublet $\Psi$ 
(with $\tilde{\Psi}\equiv -\sigma_2 \Psi^c \sigma_2 = \Psi$) having zero $B-L$ charge. 
In $2\times2$ matrix representation, the bi-doublet $\Psi$ can be
written explicitly in terms of two Dirac field $\psi^\pm$ and $\psi^0$ as~\cite{Garcia-Cely_LR_pheno}:
\begin{eqnarray}
\Psi = 
\begin{bmatrix}
\psi^0 & \psi^+ \\
\\
\psi^{-} & -\left(\psi^0\right)^c
\end{bmatrix}
\label{eq:dirac_bidoublet}.
\end{eqnarray}
\noindent The Lagrangian
for $\Psi$ is given by
\begin{eqnarray}
\mathcal{L}_{\rm BD} = \dfrac{1}{2} 
{\rm Tr}\left[\overline{\Psi}i\slashed{D}\Psi\right]-
\dfrac{1}{2}M_{\Psi} {\rm Tr}\left[\overline{\Psi}\Psi\right]\,,
\label{eq:Lagrangian_BD}
\end{eqnarray}
with covariant derivative:
\begin{eqnarray}
D_{\mu} \Psi = \partial_{\mu} \Psi - i\dfrac{g_L}{2}
\sigma_{a}{W_{L}^{a}}_{\mu} \Psi + i\dfrac{g_R}{2}
\Psi\,\sigma_{a}{W_R^{a}}_{\mu}\,.
\end{eqnarray}

Substituting, the expressions of $D_{\mu}$ and
$\Psi$ in Eq.\,\eqref{eq:Lagrangian_BD}, the Lagrangian
in terms of the component fields reads:
\begin{eqnarray}
\mathcal{L}_{\rm BD} &=& 
i\overline{\psi^0} \slashed{\partial} \psi^0 +
i\overline{\psi^{-}} \slashed{\partial} \psi^{-}+
\dfrac{g_L}{2}\left(\overline{\psi^0}\slashed{W}^3_L{\psi^0}
-\overline{\psi^{-}}\slashed{W}^3_L{\psi^{-}}
+\sqrt{2}\,\,\overline{\psi^0}\slashed{W}^{+}_L\psi^{-}
+\sqrt{2}\,\,\overline{\psi^-}\slashed{W}^{-}_L\psi^{0}\right) \nonumber \\
&&~~~~~-\dfrac{g_R}{2} \left(\overline{\psi^0}\slashed{W}^3_R{\psi^0}
+\overline{\psi^{-}}\slashed{W}^3_R{\psi^{-}}
+\sqrt{2}\,\,\overline{\psi^0}{\slashed{W}^{-}_R}\psi^{+}
+\sqrt{2}\,\,\overline{\psi^{+}}{\slashed{W}^{+}_R}\psi^{0}
\right)\nonumber\\
&&~~~~~-M_{\Psi} \overline{\psi^0}\psi^0
-M_{\Psi} \overline{\psi^{-}}\psi^{-}\,.
\label{eq:Lag_BD}
\end{eqnarray}
Although in Eq.\,\eqref{eq:Lag_BD} $\psi^{\pm}$ and $\psi^0$ have the degenerate tree level mass $M_{\Psi}$, a mass splitting
\begin{eqnarray}
M_{\psi^Q}-M_{\psi^0} &\simeq& Q\left(Q+\dfrac{2Y}{\cos\theta_W}
\right)\Delta{M}\,,
\label{eq:DeltaM}\\
&&\hspace{-6cm}{\rm with} \nonumber \\
\Delta{M} &=& \dfrac{g^2_L}{4\pi} M_{W_L} \sin^2 \frac{\theta_W}{2}
\simeq 166\,\,{\rm MeV}\,, \nonumber
\end{eqnarray}
\noindent is generated between the neutral and charged components by radiative corrections~\cite{Cirelli:2005uq}. Due to such mass splitting the charged component $\psi^\pm$ is unstable and decays
into $\psi^{0}$ and $\pi^\pm$ or a pair of light leptons on very short timescales $\mathcal{O}(10^{-11})$ s~\cite{Garcia-Cely_LR_pheno}. Notice that the specific representation of $\Psi$ in Eq.~(\ref{eq:dirac_bidoublet}) only allows the terms in Eq.~(\ref{eq:Lagrangian_BD}), automatically preventing the decay of its lightest component $\psi_0$, that is then a natural DM candidate. 

Moreover the Dirac nature of $\psi^0$ is only protected by the
$SU(2)_{L}\times SU(2)_{R}$ symmetry, which is no longer a good symmetry
after the bi-doublet $\Phi$ acquires non--vanishing VEVs.
Specifically, the $W^\pm_{R}-W^\pm_{L}$ mixing induces a transition at the one--loop level between $\psi^0$
and $(\psi^0)^c$ (with either $W^{\pm}_1$ or $W^{\pm}_{2}$
and $\psi^\mp$ running in the loop) that generates a tiny ($\sim\mathcal{O}$(keV--MeV)) off-diagonal Majorana mass term $\delta{M}$ between $\psi^0$ and $(\psi^0)^c$ proportional to $\sin(2\xi)$~\cite{Garcia-Cely_LR_pheno}. 
Consequently, the Dirac fermion splits into two quasi-degenerate Majorana fermions $\chi_{{}_{1,\,2}}$
with opposite CP given by: 
\begin{eqnarray}
\chi_{{}_{1,\,2}} &=& \dfrac{1}{\sqrt{2}}\left(\psi^0 \mp (\psi^0)^c\right)\,,\\
&&\hspace{-5cm}{\rm with} \nonumber  \\
M_{\chi_{1,\,2}} &=& M_{\Psi} \mp \delta{M}\, ,
\end{eqnarray}

\noindent providing in a natural way a specific realization of the IDM scenario that we discussed at the phenomenological level in Section~\ref{sec:capture}. The mass splitting between $\chi_1$ and $\chi_2$ is given by~\cite{Garcia-Cely_LR_pheno}:
\begin{equation}
\delta = 2 \hspace{0.6mm} \delta{M} = \frac{g_L^2}{16 \pi^2} \frac{g_R}{g_L} \sin(2\xi) M_{\Psi} 
\left[ f(r_{W_1}) - f(r_{W_2}) \right] ,
\label{eq:delta}
\end{equation}
where $r_V = M_V / M_{\Psi}$ and the loop function $f(r_V)$ given by: 
\begin{equation}
f(r_V) = 2 \int_0^1 dx \hspace{0.6mm} (1 + x) \log\left[ x^2 + (1-x) r^2_V \right] .
\label{eq:f_rV}
\end{equation}
In particular, in the phenomenological analysis of Section~\ref{sec:LR_WD_bounds} we will assume $v_1 = v_2$ in order to maximize the splitting $\delta$ at fixed $M_{W_2}$ and $g_R$ (see Eq.~(\ref{eq:tan_xi})).

Finally, in terms of the physical states $\chi_1$ and $\chi_2$
Eq.\,\eqref{eq:Lag_BD} is given by:
\begin{eqnarray}
\mathcal{L}_{\rm BD} &=& 
\dfrac{i}{2}\overline{\chi_1} \slashed{\partial} \chi_1 +
\dfrac{i}{2}\overline{\chi_2} \slashed{\partial} \chi_2 +
i\overline{\psi^{-}} \slashed{\partial} \psi^{-}
-\dfrac{M_{\chi_1}}{2} \overline{\chi_1}\chi_1
-\dfrac{M_{\chi_2}}{2} \overline{\chi_2}\chi_2
-M_{\psi^{-}} \overline{\psi^{-}}\psi^{-} \nonumber \\
&& +\dfrac{1}{2}\overline{\chi_1}
\left(g_{L} \slashed{W}^3_L - g_{R} \slashed{W}^3_R \right)\chi_{2}
- \dfrac{1}{2}\overline{\psi^{-}}
\left(g_{L} \slashed{W}^3_L + g_{R} \slashed{W}^3_R \right)\psi^{-}\nonumber\\
&& + \left\{\dfrac{1}{2} \overline{\chi_1}
\left(g_L\slashed{W}^{+}_{L} - g_R\slashed{W}^{+}_{R}\right)\psi^{-}
+ \dfrac{1}{2} \overline{\chi_2}
\left(g_L\slashed{W}^{+}_{L} + g_R\slashed{W}^{+}_{R}\right)\psi^{-}
+{\rm h.c.} \right\},
\label{eq:Lag_chi12}
\end{eqnarray}
where, $M_{\psi^{-}} \simeq M_{\Psi} + (1+\sec\theta_W)\Delta{M}$
from Eq.\,\eqref{eq:DeltaM}.

Depending on the mass splitting $\delta$ the heavier bi-doublet state $\chi_2$ decays into the lighter state $\chi_1$ via different channels, with the radiative decay $\chi_2 \rightarrow \chi_1 \gamma$ always kinematically allowed having a lifetime $\simeq 10^{-3} \hspace{0.5mm} {\rm s} 
\left( \frac{1 \hspace{0.5mm} {\rm MeV}}{\delta} \right)^3$~\cite{Garcia-Cely_LR_pheno} (such expression is valid when 
$M_\Psi \simeq  {\cal O}$(TeV) and $\delta\ll M_\Psi$).
As a consequence, for the choice of parameters in the next Section, by the present time all the $\chi_2$ states 
have decayed and $\chi_1$ is the DM candidate in our Universe.

\subsection{Bounds from White Dwarves on LRSM Bi--doublet Dark Matter }
\label{sec:LR_WD_bounds}
In this Section we will identify $\chi$ and $\chi^{\prime}$ with 
$\chi_1$ and $\chi_2$. The process of inelastic scattering of $\chi$ against a WD nucleus occurs dominantly through the exchange of the light $Z$-boson. The corresponding interaction Lagrangian (involving $\chi$, $\chi^{\prime}$ and $Z$) can be written as:
\begin{equation}
\mathcal{L}_{\chi \chi^{\prime} Z} = g_{\chi} \hspace{0.5mm} \overline{\chi} \slashed{Z} \chi^{\prime} ,
\label{eq:Lag_chi_Z}
\end{equation}
where
\begin{equation}
g_{\chi} = \frac{1}{2} \left( g_L \cos\theta_W \cos\phi + g_R \sin\theta_W \sin\theta_R \cos\phi + g_R \cos\theta_R \sin\phi \right) .
\label{eq:g_chi}
\end{equation}
The coupling $g_{\chi}$ is obtained by expressing $W^3_L$ and $W^3_R$ in Eq.~(\ref{eq:Lag_chi12}) in terms of the mass eigenstate $Z$, 
using the neutral gauge boson mixing matrix given 
in the appendix of~\cite{Garcia-Cely_LR_pheno}.
Since the typical values of the momentum transfer in such a process are always much lower than $M_{Z}$, one can express the interaction of $\chi$ and $\chi^{\prime}$ with SM quarks in terms of the following dimension-6 operators:
\begin{equation}
\mathcal{O}^{(6)}_{V,q} = C^{(6)}_{V,q}\left( \overline{\chi} \gamma_\mu \chi^{\prime} \right)
\left( \overline{q} \gamma^\mu q \right) , \hspace{4mm}
\mathcal{O}^{(6)}_{A,q} =C^{(6)}_{A,q}\left( \overline{\chi} \gamma_\mu \chi^{\prime} \right)
\left( \overline{q} \gamma^\mu \gamma_5 q \right) ,
\label{eq:Lag_chi_q}
\end{equation}

\noindent with:

\begin{equation}
    C^{(6)}_{V,q}=\frac{g_{\chi} g_L g^q_V}{M^2_{Z} \cos\theta_W}, \hspace{4mm}C^{(6)}_{A,q}=\frac{g_{\chi} g_L g^q_A}{M^2_{Z} \cos\theta_W}\,, 
\end{equation}

\noindent where $g^q_V$ and $g^q_A$ are defined in~\cite{SM_couplings_Romao2012}. 

In the non--relativistic limit the above interaction terms lead to a WIMP--nucleon interaction driven by the effective Hamiltonian:

\begin{equation}
    {\cal H}=\sum_{\mathcal{N}=p,n} ( c_1^{\mathcal{N}} {\cal O}_1^{\mathcal{N}} + c_7^{\mathcal{N}} {\cal O}_7^{\mathcal{N}} + c_9^{\mathcal{N}} {\cal O}_9^{\mathcal{N}}), 
\end{equation}
\noindent (in the operator base of~\cite{haxton1}), with $p,n$ indicating proton and neutron. The Wilson coefficients $c_1^{\mathcal{N}}$, $c_7^{\mathcal{N}}$ and $c_9^{\mathcal{N}}$ in terms of the $C^{(6)}_{V,q}$ and $C^{(6)}_{A,q}$ quantities can be obtained for instance, from~\cite{form_factor_Bishara2017}. In particular,  the effective operator ${\cal O}_1$ leads to a spin--independent WIMP--nucleus cross-section  proportional to the square of the atomic mass number of the target, while the ${\cal O}_7$ and ${\cal O}_9$ operators, which are velocity and momentum suppressed, vanish in the case of the spinless targets ($^{12}C$, $^{16}O$ or $^{20}Ne$) that constitute the interior of a WD. Specifically, for the operator $\mathcal{O}^{(6)}_{V,q}$ one has $c_1^{\mathcal{N}}$ = $\sum_{q=u,d} C^{(6)}_{V,q} F_{\mathcal{N}}^{(q)}$ with:

\begin{eqnarray}
F^{(u)}_p&=&2,\hspace{0.5mm}F^{(d)}_p=1,\\
F^{(u)}_n&=&1,\hspace{0.5mm}F^{(d)}_n=2,
\end{eqnarray}
    
\noindent while the WIMP--nucleon cross-section $\sigma_{\mathcal{N}}$ introduced in Section~\ref{sec:capture} is given by 
$\sigma_{\mathcal{N}}=(c_1^{\mathcal{N}})^2\mu_{\chi \mathcal{N}}^2/\pi$. 
In order to evaluate the WD capture rate we have assumed that the WDs observed in M4 are made of carbon and we have calculated the differential cross-section  $\frac{d\sigma[\chi+N \rightarrow \chi^{\prime}+N]}{dE}$ in Eq.~(\ref{eq:interaction_rate}) using the \verb|WimPyDD| code~\cite{wimpydd,wimpydd_conf}.

\begin{figure*}[ht!]
\centering
\includegraphics[width=7.49cm,height=7.4cm]{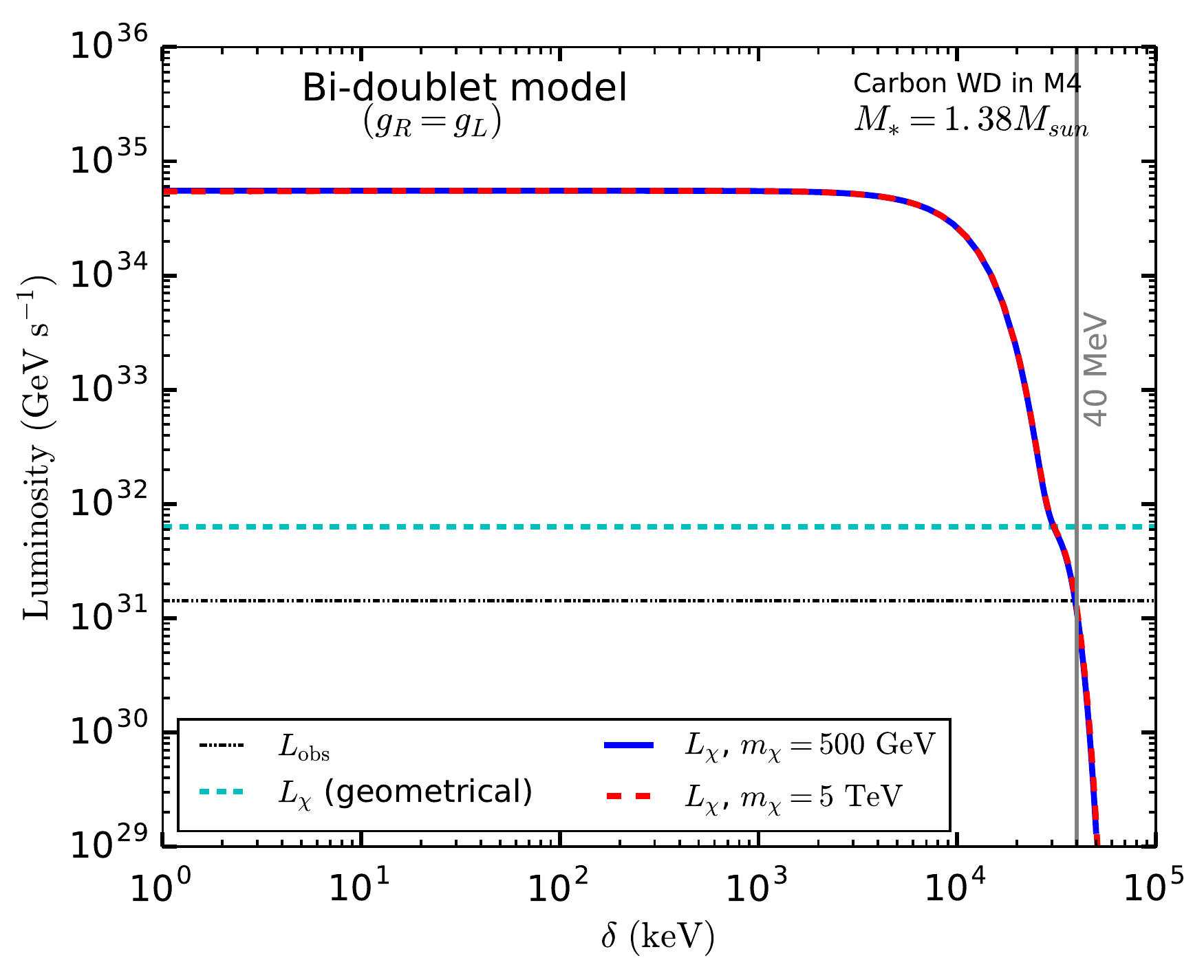}
\includegraphics[width=7.49cm,height=7.4cm]{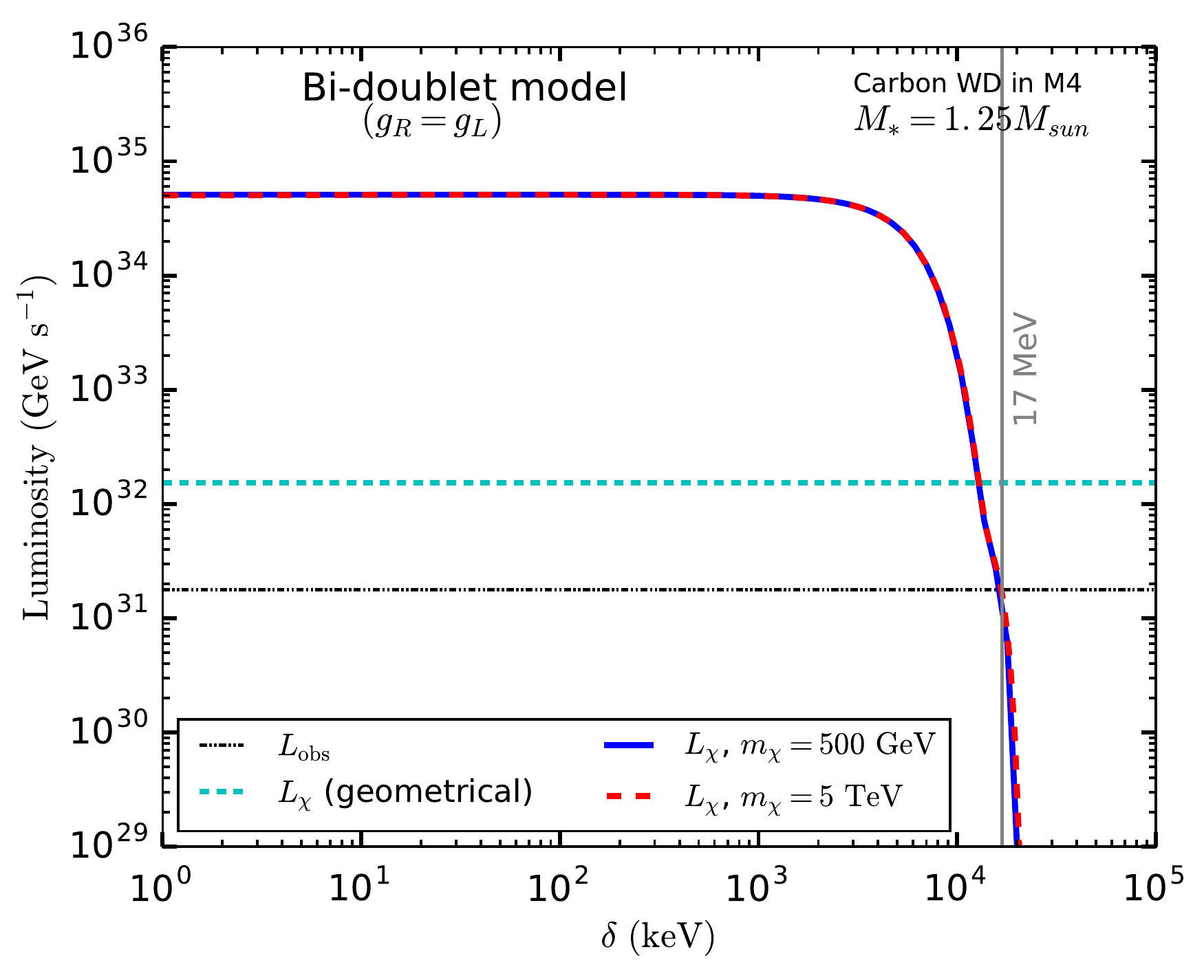}
\caption{Bi-doublet DM induced luminosities in optically thin (solid blue and red dashed lines) and geometric (horizontal cyan lines) limits as functions of the mass splitting $\delta$, considering two benchmark carbon WDs in M4 
with masses $M_* = 1.38$ \(M_\odot\) (left panel) and $M_* = 1.25$ \(M_\odot\) (right panel). 
The solid blue line and the red dashed line correspond to the capture of 500 GeV and 5 TeV DM particles, respectively. 
The black dotted lines in both panels represent the luminosities observed for 
the considered WDs~\cite{improved_WD_2021}. The vertical gray lines indicate the maximum values of $\delta$ that are possible to probe for these WDs.}
\label{fig:cap_Lumi_BD}
\end{figure*}

In Fig.~\ref{fig:cap_Lumi_BD} the solid blue line and the dashed red line represent the result of the evaluation of the expected luminosity of Eq.~(\ref{eq:L_chi})
for the capture of LRSM bi--doublet DM by WDs in the optically--thin limit as a function of the mass splitting parameter $\delta$ for DM masses $m_\chi$ = 500 GeV and $m_\chi$ = 5 TeV, respectively, for $g_L=g_R$ and for the the same carbon WDs of Fig.~\ref{fig:cap_Lumi_SI}. In both plots the cyan and black dotted horizontal lines are the same of Fig.~\ref{fig:cap_Lumi_SI} and represent, respectively, the geometrical limit of the capture rate and the corresponding observed luminosities~\cite{improved_WD_2021}. As expected, unless the capture process in the WD is kinematically forbidden the large WIMP--nucleus cross-section implied by the non--vanishing hypercharge of the bi--doublet saturates the geometrical capture upper bound and exceeds the observation. This is achieved for $\delta\gsim$ 40 MeV (17 MeV) for $M_* = 1.38$ \(M_\odot\) 
($M_* = 1.25$ \(M_\odot\)).

Notice that, as described in Section~\ref{sec:model_description}, 
the heavier bi-doublet state $\chi^{\prime}$ decays to the DM state $\chi$ 
within a time-scale that is extremely small compared to the typical age of the WDs in M4, 
expected to be a few Gyrs. In such a case, the expressions of the capture rate given in Section~\ref{sec:capture_in_wd} apply, and the final population of DM particles in the WD core consists of only $\chi$. The pair-annihilation of these particles, which is driven by t-channel processes, injects electroweak gauge bosons whose cascade products thermalise in the WD medium and heat up the star.

\begin{figure*}[ht!]
\centering
\includegraphics[width=7.49cm,height=7.4cm]{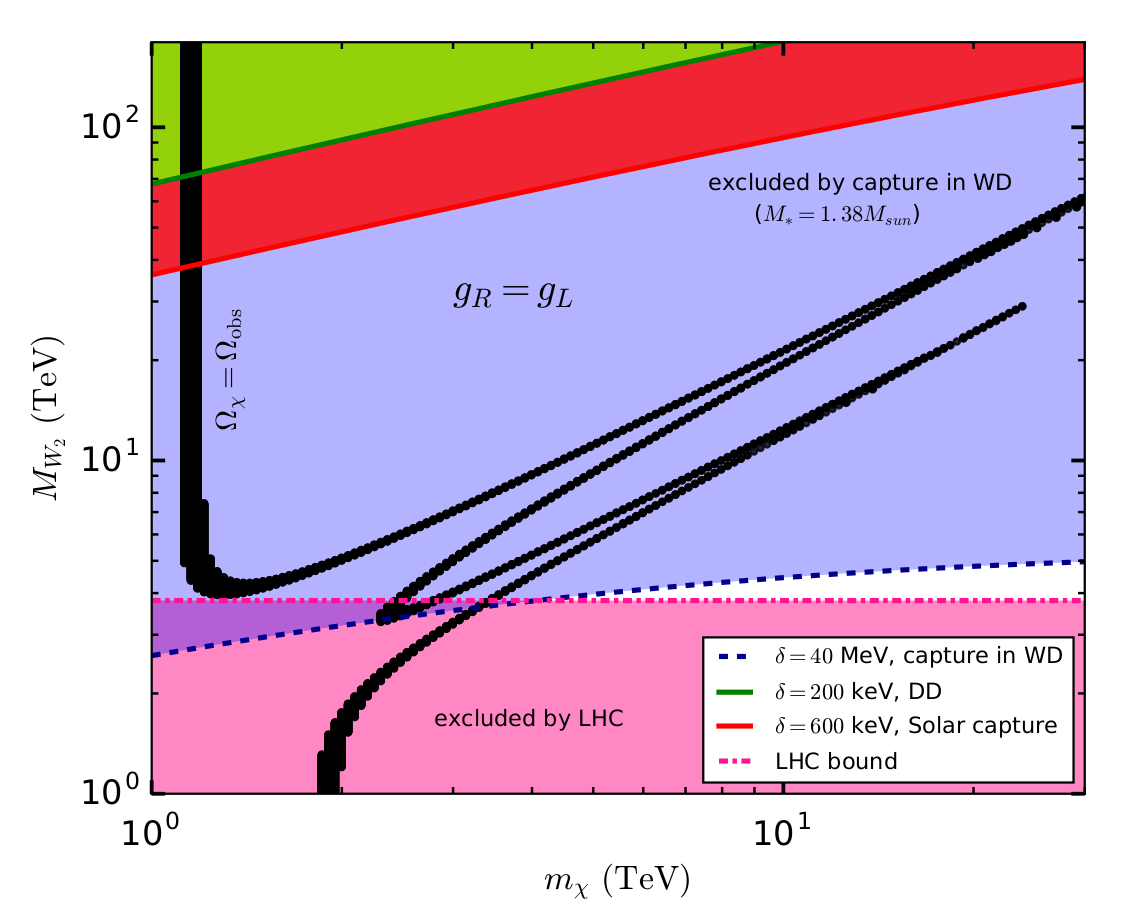}
\caption{The region of the bi-doublet parameter space (spanned by $M_{W_2}$ and the bi-doublet DM mass $m_{\chi}$) that are excluded by the observation of the faintest and heaviest ($M_* = 1.38$ \(M_\odot\)) WD (assuming it is made of carbon) in M4 is shown by the blue shaded area, 
considering $g_R = g_L$. The pink, green and red shaded areas correspond to the regions 
excluded by LHC search, direct detection (DD) experiments and solar neutrino search, respectively. 
The black band consists of parameter points where the calculated thermal relic density is in the observed range.}
\label{fig:MW2_Mx}
\end{figure*}

\begin{figure*}[ht!]
\centering
\includegraphics[width=7.49cm,height=7.4cm]{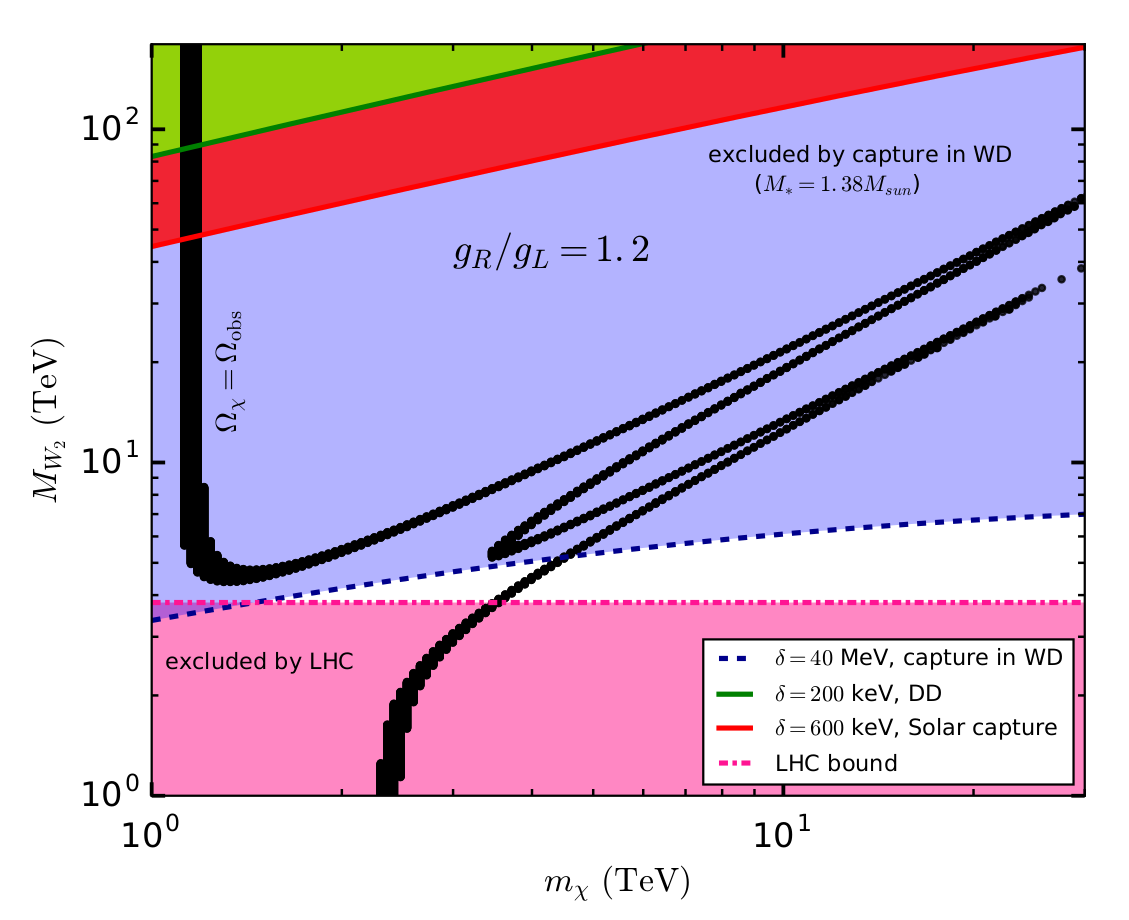}
\includegraphics[width=7.49cm,height=7.4cm]{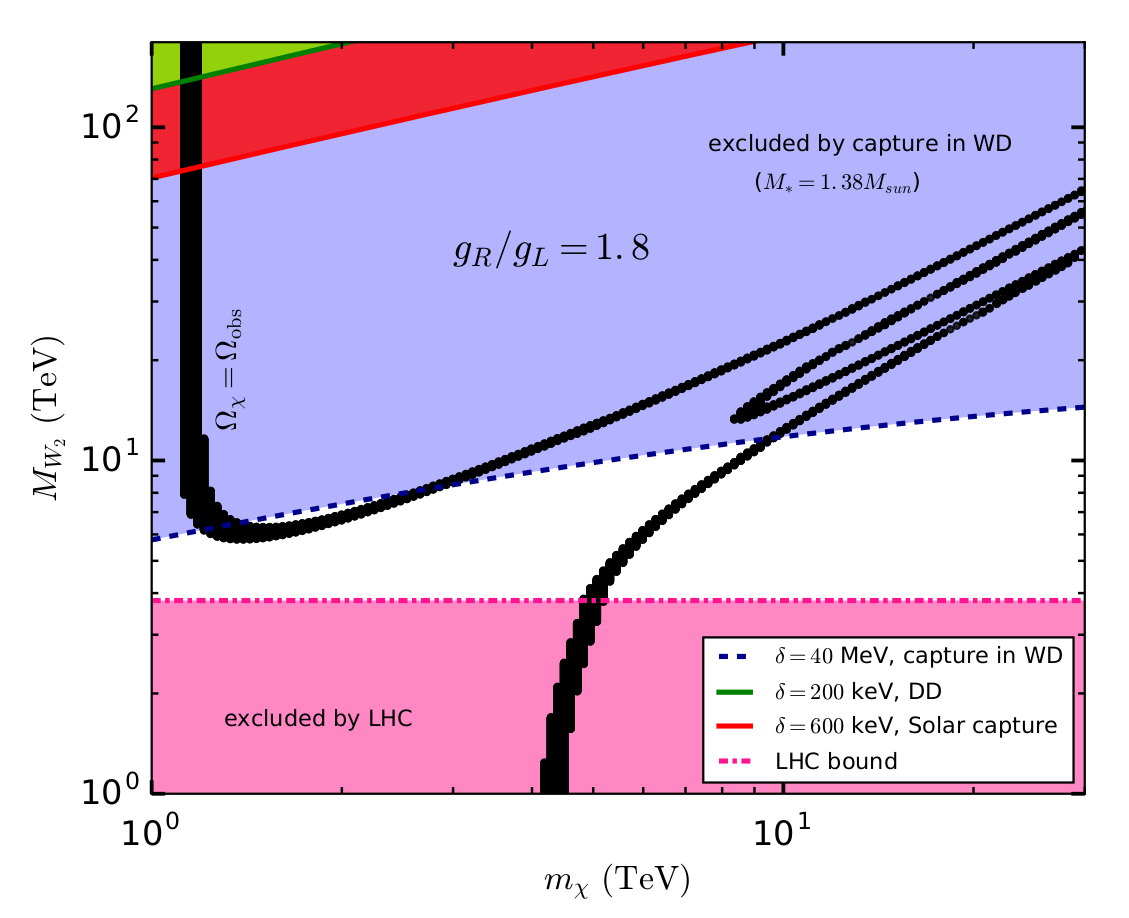}
\caption{Same as Fig.~\ref{fig:MW2_Mx}, but considering scenarios with 
$g_R / g_L = 1.2$ (left panel) and $g_R / g_L = 1.8$ (right panel).}
\label{fig:MW2_Mx_2}
\end{figure*}

As already pointed out in Section~\ref{sec:capture}, the bound from WDs is much more constraining than that expected from direct detection and from solar neutrino searches. The impact from such bound on the $m_\chi$--$M_{W_2}$ parameter space of the LRSM bi-doublet Dark Matter is shown in Fig.~\ref{fig:MW2_Mx} for $g_R = g_L$, which represents the most commonly used value in phenomenological and experimental analyses. In the upper--left corner the green--shaded area shows the part of parameter space excluded by direct detection, which can probe $\delta\lsim$ 200 keV~\cite{iDM_DD_Nagata2014}, while the red--shaded one the part of parameter space excluded by the solar neutrino searches, which can probe $\delta\lsim$ 600 keV~\cite{idm_sun_Catena_2018}. On the other hand, the pink--shaded region is excluded by the non observation of the $W_2$ boson at the LHC ($M_{W_2}<$ 3.8 TeV~\cite{lch_mw2}), and the black band represents the region where the thermal relic density is in the observed range (0.11$\le \Omega h^2\le$0.13). As shown in such figure, when only the direct detection, solar neutrino searches and the LHC bounds are considered (as done in the literature so far~\cite{Garcia-Cely_LR_pheno}) the LRSM bi--doublet is a viable DM candidate in a wide range of masses, 1.1 TeV $\lsim m_\chi\lsim$ 30 TeV and for $g_R = g_L$. However, when the bound from WD is considered this changes considerably: in Fig.~\ref{fig:MW2_Mx} the blue shaded area, that lies above the dashed blue line for $\delta$ = 40 MeV, represents the parameter space excluded by WDs. Indeed, for $g_R = g_L$ the full cosmologically viable parameter space now lies in such region and is ruled out. As pointed out below Eq.~(\ref{eq:delta}) in our calculation we have maximized $\delta$ at fixed $m_\chi$, $g_R$ and $M_{W_2}$ in terms of the other parameters of the model. As a consequence, for other choices of such parameters the $M_{W_2}$ values excluded by WDs would be even lower, so that the excluded region of Fig.~\ref{fig:MW2_Mx} should be considered as a conservative one. 

In order to evade the WD constraint one needs to increase further the value of the mass splitting parameter $\delta$ at fixed $m_\chi$ and $M_{W_2}$. As shown in Eq.~(\ref{eq:delta}), the only way to achieve this within the available parameter space is to chose $g_R>g_L$. 
This is done in Fig.~\ref{fig:MW2_Mx_2}, where $g_R/g_L$ =1.2 in the left--hand plot, and $g_R/g_L$ =1.8 in the right--hand plot. Indeed, now in both cases some cosmologically viable parameter space is recovered. In particular, for $g_R/g_L$ =1.2 this requires $M_{W_2}\lsim$ 6 TeV, and 3.5 TeV $\lsim m_\chi\lsim$ 5 TeV, while for $g_R/g_L$ =1.8 $M_{W_2}\lsim$ 12 TeV and two bi-doublet allowed mass ranges are found: 1.2 TeV $\lsim m_\chi\lsim$ 3 TeV and 5 TeV $\lsim m_\chi\lsim$ 10 TeV. In Section~\ref{sec:high_scale} we will discuss how $g_R/g_L>1$ can be achieved at low energy when the LRSM model is embedded at high scale in a Grand Unified Theory, and provide an explicit example with $g_R/g_L\simeq$ 1.2. On the other hand we consider $g_R/g_L$ =1.8 a reasonable estimation of the maximal allowed value of such ratio since it roughly coincides to the perturbative upper bound discussed in~\cite{gr_gl_perturbativity}\footnote{See Fig. 7 of~\cite{gr_gl_perturbativity}. However, at variance with Fig.~\ref{fig:MW2_Mx_2}, in the discussion of Ref.~\cite{gr_gl_perturbativity} $g_R/g_L$=1.8 is correlated to a high $M_{W_2}$ mass, $M_{W_2}\gsim$ 90 TeV.}. 

For the calculation of the bi--doublet relic abundance in Figs.~\ref{fig:MW2_Mx} and \ref{fig:MW2_Mx_2}  we have modified the \verb|FeynRules|~\cite{feynrules} implementation of the LRSM model of Ref.~\cite{LR_feynrules} in order to allow $g_R\ne g_L$, and used it to produce a \verb|CalcHEP|~\cite{calchep} output file, that we have then modified implementing the Lagrangian of Eq.~(\ref{eq:Lag_BD})~\footnote{The relic abundance can be directly calculated in terms of the $\Psi^0$ and $\psi^{\pm}$ Dirac fermions of the bi-doublet of Eq.~(\ref{eq:dirac_bidoublet}), since in the early Universe the effect of the mass splitting is negligible in the calculation of the annihilation cross-section.}. Finally, we have used such modified \verb|CalcHEP| file in \verb|MicrOMEGAs|~\cite{micromegas}. 

\subsection{Consequences at High Scale \label{sec:high_scale}} 

The implementation of the bound from WDs in the phenomenological analysis of Section~\ref{sec:LR_WD_bounds} singles out a relatively well defined parameter space in order for the LRSM bi--doublet scenario to provide a viable DM candidate, characterized by a low mass for the $W_2$ boson ($M_{W_2}\lsim$ 10 TeV) and $g_R>g_L$. In this Section we wish to discuss whether such parameter space can be obtained, and how naturally, when  
the LRSM model is embedded in a Grand Unified Theory, which represents its most popular ultraviolet completion.

The LRSM scenario naturally arises in Grand Unified Theories (GUT). Indeed, many scenarios have been discussed in the literature \cite{Pati:1974yy,Kibble:1982dd,Rizzo:1981jr,Rizzo:1981dm,Mohapatra:1980yp,Deshpande:1990ip}, where the LR group $SU(2)_L\times SU(2)_R$ is embedded into a GUT SO(10). Specific examples include: (i) SO(10)$\supset SU(2)_L\times SU(2)_R \times SU(4)_C \times {\rm D}$; (ii) SO(10)$\supset SU(2)_L\times SU(2)_R \times SU(4)_C$,
(iii) SO(10)$\supset 
{\rm LRSMD}\equiv SU(2)_L\times SU(2)_R \times SU(3)_C \times U(1)_{B-L} \times {\rm D}$  and
(iv) SO(10)$\supset {\rm LRSM}\cancel{\rm D}\equiv SU(2)_L\times SU(2)_R \times SU(3)_C \times U(1)_{B-L}$. Here D stands for the parity symmetry that leaves the theory 
invariant by the interchange of any multiplet of $SU(2)_L$ into the corresponding  $SU(2)_R$ multiplet. 
The low--energy scenario discussed in Sections~\ref{sec:model_description} and \ref{sec:LR_WD_bounds} 
corresponds to either LRSMD or LRSM$\cancel{\rm D}$ extended with the DM bi--doublet $\Psi$ 
(LRSMD + $\Psi$ and   
LRSM$\cancel{\rm D}$ + $\Psi$ in the following).

In the analysis of Section~\ref{sec:LR_WD_bounds} we have used WDs to constrain the bi--doublet LRSM parameter space at the phenomenological level, i.e. taking the low--energy parameters of the model introduced in Section~\ref{sec:model_description} ($M_{\chi_1}$ = $m_{\chi}$, $M_{W_2}$, $g_R$ and $v_1/v_2$) as independent without assuming any specific ultraviolet completion. In a GUT theory context such parameters are expected to be correlated. In the literature most often the so called \emph{minimal} version of the LRSM assumes D parity.
However, in the following we will consider both cases, conserved D parity and broken D parity. 

The most important appeal of GUT is the unification of gauge couplings at high-energy, which comes together with the salient prediction of proton decay. Therefore, once the embedding of the LR models in a GUT framework comes into play we need to pay attention to the GUT scale and to the proton decay rate.  In particular, in some of its realizations LRSM models embedded in SO(10) typically achieve a unification scale around $10^{15}$ GeV, which leads to proton decay rates incompatible to the experimental constraints. Specifically, for the minimal particle content of \Tabref{tb:LRSM_MC} and the embedding (i) the breaking of the $SU(2)_R$ is achieved at around $\MR\approx 10^{13}$ GeV and the GUT scale is
$\MG\approx 10^{15}$ GeV. On the other hand, for (ii) $\MR$ is around $10^{11}$ GeV, while $\MG\approx 10^{16}$ GeV. For (iii) LRSMD,  $\MR\approx 10^{10}$ GeV, while the unification scale is achieved around $\MG \approx 10^{15}$ GeV. Finally for (iv) {\rm LRSM}\cancel{\rm D}, $\MR \approx 10^9$ GeV and $\MG \approx 10^{16}$ GeV~\cite{Shaban:1992he}~\footnote{For the exact values of $\MR$ and $\MG$ in each case check the original references and the updates of \cite{Mambrini:2015vna,Chakrabortty:2017mgi}}. 

These basic scenarios hence do not offer an explanation of why $v_R\simeq M_{W_2} \lsim 10^{4}$ GeV, as emerging from Fig.~\ref{fig:MW2_Mx} (notice that, depending on the specific mechanism of symmetry breaking, it is natural to expect that $v_R$ also determines the WIMP mass $m_\chi$).

\begin{figure*}[ht!]
\centering
\includegraphics[width=7.49cm,height=7.4cm]{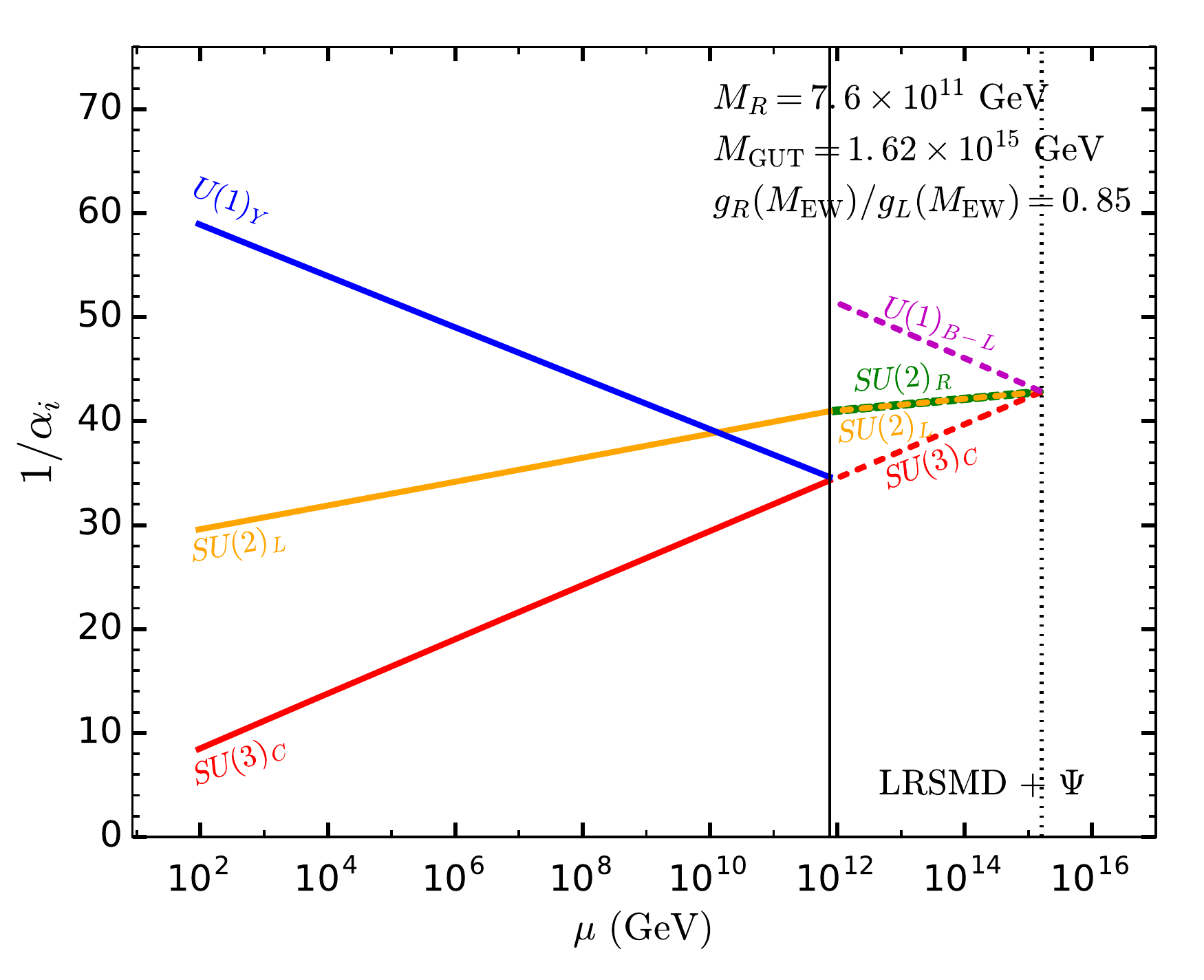}
\includegraphics[width=7.49cm,height=7.4cm]{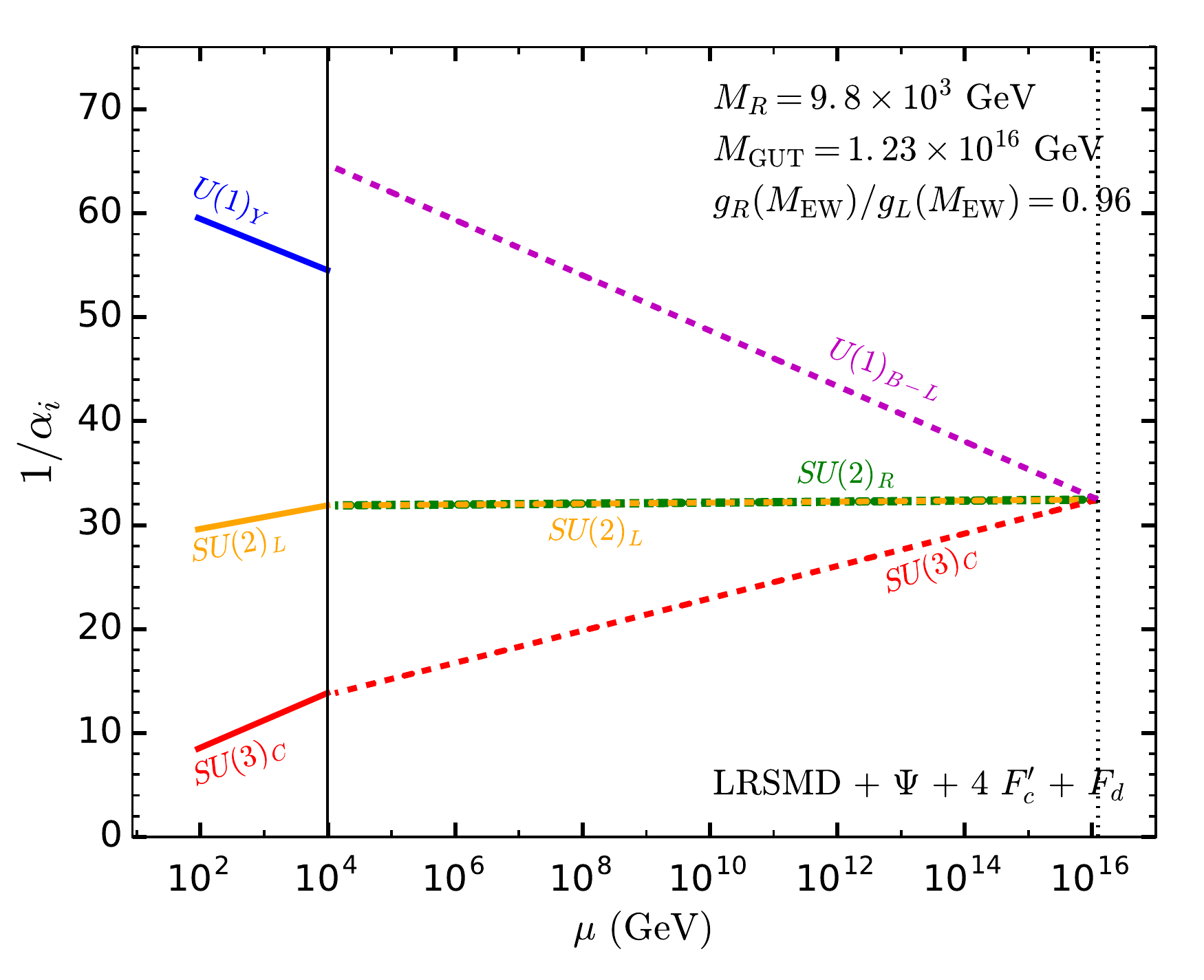}
\includegraphics[width=7.49cm,height=7.4cm]{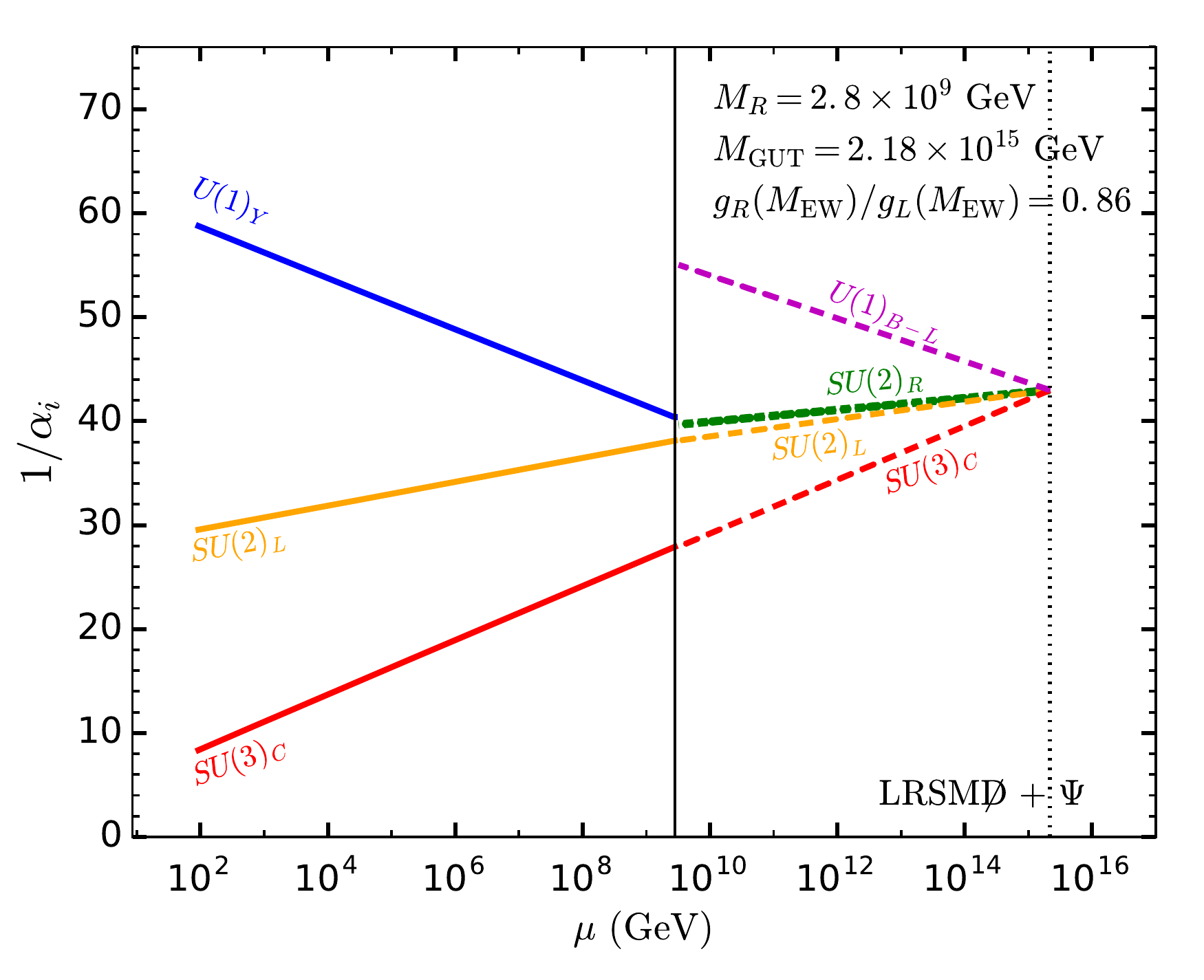}
\includegraphics[width=7.49cm,height=7.4cm]{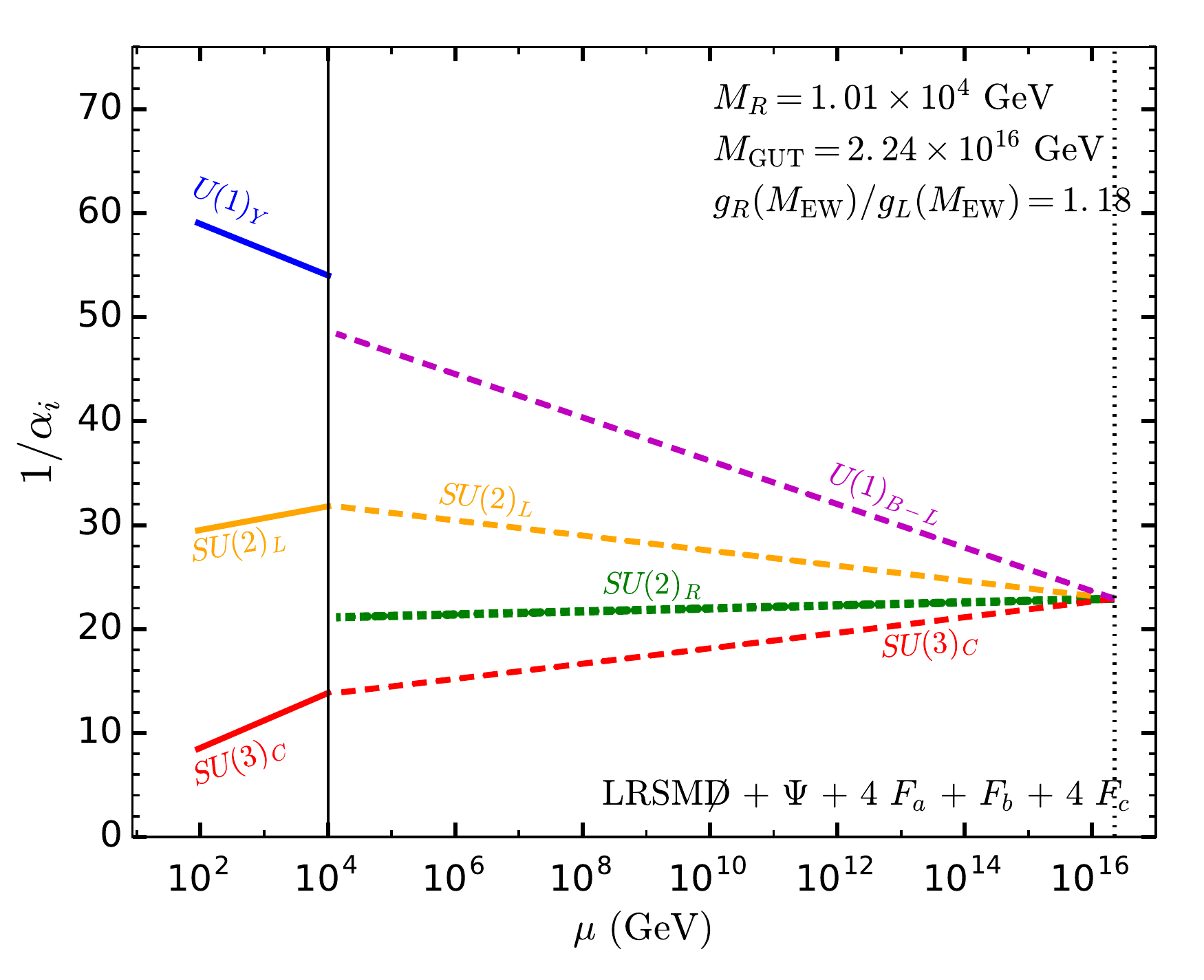}
\caption{Running of the inverse of the gauge couplings $\alpha_i$ of the Left Right Symmetric models we consider. Top panels: D parity conserving scenarios. Top-Left: LRSMD  $+\, \Psi$, Minimal Model plus Dark Matter bi-doublet. Top-Right: LRSMD  $+\,  \Psi \, + 4\, F_c^\prime  + F_d$.
Bottom panels: scenarios with broken D parity.
Bottom-Left: LRSM\cancel{D}  $+\, \Psi$, Minimal Model plus Dark Matter bi-doublet.  
Bottom-Right: LRSM\cancel{D}  $+\, \Psi\, + 4\, F_a + F_b + 4\, F_c$. The definition of the fermions $F_i$ is given in Tab.~\ref{tb:LRSM_MC}. 
The value of $\MR$ ($M_{\rm GUT}$) corresponding to each case is indicated by the black solid (dotted) line.}
\label{fig:running_models}
\end{figure*}

Extending the LRSMD model by adding only the DM bi--doublet fermion (the LRSMD + $\Psi$ case) actually slightly increases the value $\MR\simeq v_R$ of the scale at which $SU(2)_R$ is broken and increases the proton decay (PD) lifetime (see \Tabref{tb:LRSM_PD})~\footnote{It is well known that in non-supersymmetric theories the dominant decay channel is $p\rightarrow \pi^0 e^+$. For specific formulas see for example Appendix D of \cite{Mambrini:2015vna} and Sections 3 and 4.2 of \cite{Chun:2021brv}.}. So although adding $\Psi$ to the model can increase the PD lifetime (albeit not enough in most cases) it is of no help to reduce $\MR$. This case is shown in the top left plot of \Figref{fig:running_models}, where the running of the coupling constants is plotted as a function of the energy scale $\mu$. As a consequence, in order to lower $\MR$ one needs to add extra-matter (in particular, all the additional degrees of freedom discussed in the following examples are listed in Tab.~\ref{tb:LRSM_MC}). In~\cite{Lindner:1996tf} it was found that adding color multiplets is  key to lower $\MR$ down to $10^5$ GeV or $10^4$ GeV. Notice that such additional high--scale degrees of freedom do not affect the phenomenology discussed in Section~\ref{sec:LR_WD_bounds}. In the top-right plot of Fig.~\ref{fig:running_models} we show a specific example of this, where the MLRSM model is extended to include, besides the DM fermion bi-doublet $\Psi$, four fermions  $F_c'$, which are triplets under $SU(3)_C$ and singlets under all other interactions except for the D parity charge, and one fermion $F_d$, which is a bi-doublet under $SU(2)_R$ and $SU(2)_L$ and can have a non-zero D parity charge. We can see that now the scale $\MR$ is lowered. However, the ratio $g_R/g_L$ is still smaller than one at the electroweak scale $M_{\rm EW}\simeq 10^2$ GeV. 

The feature $g_R(\MEW)/g_L(\MEW)<1$ (that for simplicity will be just indicated as $g_R/g_L$ in the discussion below) is always present in D parity conserving models since the running of $g_R$ and $g_L$ only splits below $M_R$, where $g_R$ practically stops evolving while the value of $g_L$ gets larger as the energy scale is decreased (notice that in Fig.~\ref{fig:running_models} inverse couplings are plotted). 
So the only way to obtain $g_R/g_L>$1 is to break D parity at high scale, i.e. to split the running of $g_R$ and $g_L$ for $M_R<\mu<M_{\rm GUT}$, so that $g_R>g_L$ already at $M_R$. This can be achieved in different ways which are anomaly--free~\cite{Lindner:1996tf}. The most straightforward is by simply omitting the left-handed triplet $T_L$ \cite{gr_gl_perturbativity} (also this does not affect the phenomenology).
This case, indicated by LRSM$\cancel{\rm D}$ + $\Psi$, is shown in the left--bottom plot of \Figref{fig:running_models}. Since the scale $\MR$ is lower than for the LRSMD + $\Psi$ case the ratio $g_R/g_L$ turns out to be slightly bigger but nevertheless less than 1 (notice also that in this case PD is not consistent with observation). For this reason, in the final example shown in the bottom--right plot of Fig.~\ref{fig:running_models} we add more fermions transforming under $SU(2)_L$ only than under $SU(2)_R$, so that the slope of $g_L$ with respect to the energy scale $\mu$ gets steeper compared to $g_R$. In such scenario, indicated with 
LRSM$\cancel{\rm D}$ + $\Psi$ + 4 $F_a$ + $F_b$ + 4 $F_c$,  we add four fermions $F_a$ which are doublets of $SU(2)_L$, triplets of $SU(3)_C$ and singlets of the other groups, one fermion $F_b$, which is a doublet of $SU(2)_R$, triplet of $SU(3)_C$ and singlet of the other groups, and four fermions $F_c$ which are triplets of $SU(3)_c$ and singlets under the rest of the groups (see Tab.~\ref{tb:LRSM_MC}). Indeed, now $g_R/g_L\simeq$1.2 is achieved. This case then provides an example where the value of $\MR$ can be explained to be around $10^4$ GeV without any fine tuning,  $M_{\rm GUT}\gsim 10^{16}$ GeV so that the PD lifetime can be acceptable (see \Tabref{tb:LRSM_PD}), and $g_R/g_L>$1. 

The values of the ratio $g_R/g_L$ are provided for all the models discussed above in \Figref{fig:running_models}. 
To get the plots of Fig.~\ref{fig:running_models} we obtained the corresponding beta functions, given in Tab.~\ref{tab:beta_functions}, using the code \verb|pyr@te|~\cite{pyr@te} and used them in a private Runge-Kutta code.

\begin{table}[h]
\caption{
\label{tb:LRSM_PD} Proton decay lifetime $\tau(p\rightarrow \pi^0\, e^+)$[yrs]. The 1 $\sigma$ errors are extracted from a chi--square  with the observables at the electroweak scale.}\vspace*{0.1cm}
\renewcommand{\arraystretch}{1.0}
\centering
\begin{tabular}{|l@{\hskip .66cm}c|}
\hline
\hline
 \multicolumn{2}{c}{
Model Predictions }\\
\hline
LRSMD  & $ (1.3\pm 0.9 ) \times 10^{32}$ \\
LRSMD + $\Psi$ & $ (2.9\pm 1.9 ) \times 10^{32}$\\
LRSMD +  $\Psi$ + 4 $F_c^\prime$ +  $F_d$  & $(5.4 \pm 3.7) \times 10^{36}$ \\
LRSM\cancel{D} & $ (3.8\pm 2.5 ) \times 10^{36}$ \\
LRSM\cancel{D} + $\Psi$ & $(7.6\pm 5.2) \times 10^{32}$\\
LRSM\cancel{D} + $\Psi$ + 4 $F_a$ + $F_b$ + 4 $F_c$  & $(2.0 \pm 1.3) \times 10^{37}$ \\
\hline
\multicolumn{2}{c}{ Experimental bounds/discovery}\\
\hline
Current Bound: $1.6 \times 10^{34}$ at $95\%$ C.L \cite{Aoki:2017puj, Super-Kamiokande:2016exg}
& Projected Discovery: $6.3 \times 10^{34}$ \cite{Hyper-Kamiokande:2018ofw} \\
Projected Bound: $7.8 \times 10^{34}$ $90\%$ C.L \cite{Hyper-Kamiokande:2018ofw}  &  \\
\hline
\end{tabular}
\end{table}

\section{Conclusions}
\label{sec:conclusions}

Models of structure formation suggest that the inner part of globular clusters may have a DM density that largely exceeds that estimated in the solar neighborhood. Assuming this is so and that the DM is a WIMP particle, compact stars such as white dwarves can capture WIMPs through a process similar to that studied in the case of the Sun or the Earth, leading to an increase of the star luminosity through their annihilation process.  As a consequence the issue of the amount of DM in globular clusters, and specifically in M4, has drawn increasing interest in the literature~\cite{idm_wd_2010, dm_cap_WD_Dasgupta2019, improved_WD_2021}.  We have shown that if the WIMP interacts with the nuclear targets within the WD through inelastic scattering the data on low--temperature large-mass WDs in the Messier 4 globular cluster can lead to a constraint on the mass splitting $\delta\lsim$ few tens of MeV that largely exceeds that ensuing from Direct Detection in terrestrial underground experiments, $\delta\lsim$ 200 keV, and from solar neutrino searches, $\delta\lsim$ 600 keV, for a generic WIMP with a spin--independent cross-section off nucleons larger than $\simeq$ a few $\times$ 10$^{-43}$ cm$^2$, assuming conservatively that the WD is entirely composed of carbon.

 We have applied such improved constraint to the specific DM scenario of a self--conjugate bi-doublet in the Left--Right Symmetric Model (LRSM), where the standard $SU(2)_L$ group with coupling $g_L$ is extended by an additional $SU(2)_R$ with coupling $g_R$ in order to explain maximal parity violation in Weak Interactions. The ensuing bounds significantly reduce the cosmologically viable parameter space of such scenario, in particular requiring $g_R>g_L$. For instance, for $g_R/g_L$ = 1.8 we have found the two viable mass ranges for the bi--doublet DM candidate, 1.2 TeV $\lsim m_\chi\lsim$ 3 TeV and 5 TeV $\lsim m_\chi\lsim$ 10 TeV, when the charged $SU(2)_R$ gauge boson mass $M_{W_2}$ is less than $\simeq$ 12 TeV. 
 
 In Section~\ref{sec:high_scale} we have provided a short discussion on how such phenomenological parameter space at low energy can be achieved when the LRSM model is embedded at high scale in a Grand Unified Theory (GUT), showing that in the minimal scenario of LRSM + DM bi--doublet one naturally gets $M_{W_2}$, $m_\chi\gsim$ 10$^{10}$ GeV, and $g_R<g_L$, with a GUT scale $M_{\rm GUT}\simeq 10^{15}$ GeV incompatible to proton decay bounds. As a consequence,  one needs to extend the ultraviolet completion of the LRSM model with additional multiplets besides the DM bi--doublet in order for the latter to explain the DM in the Universe in agreement with the WD bounds discussed in Section~\ref{sec:LR_WD_bounds}. Specifically, we found that adding color triplets that are singlets under all other groups and carry zero B-L charges can naturally yield $M_{W_2}, m_\chi\lsim$ 10 TeV. On the other hand, the only way to obtain $g_R>g_L$ is to split the running of $g_R$ and $g_L$ by assuming that D parity is broken at high scale, i.e. by introducing an asymmetry between the multiplets of $SU(2)_L$ and $SU(2)_R$. We provided a specific example of such scenario in Fig.~\ref{fig:running_models}, with the particle content of Tab.~\ref{tb:LRSM_MC}.

We conclude by reminding that the amount of DM in M4 is still controversial. Our analyses motivates further studies to settle this issue. Moreover, at the end of Section~\ref{sec:capture_in_wd} we pointed out that the WDs bounds discussed in the present paper could be extended using more dense stellar object such as neutron stars, probing values of the inelastic splitting $\delta$ as large as $\sim 300$ MeV. In particular, a future observation of neutron stars with temperatures $T\lsim$ a few thousand Kelvin would rule out the full parameter space of LRSM bi--doublet DM.

\section*{Acknowledgements}
We thank Alexander Pukhov and Julian Heeck for useful discussions. 
We thank Lohan Sartore for helping us with running \verb|pyr@te|. 
This research was supported by the National Research Foundation of Korea(NRF) funded by the Ministry of Education through the Center for Quantum Space Time (CQUeST) with grant number 2020R1A6A1A03047877 and by the Ministry of Science and ICT with grant number 2021R1F1A1057119.  L. V.-S. acknowledges the KIAS  warm hospitality and stimulating environment in these challenging times.

\appendix
\section{Ultraviolet Completions of the LRSM Model and Beta Functions
\label{app:beta_functions}}
\begin{table}[h]
\caption{The matter content of the Minimal Left Right Symmetric Model and its extensions discussed in Section~\ref{sec:high_scale}. The LRSM with broken D parity does not contain $T_L$, while the one with D parity conservation does.
\label{tb:LRSM_MC} }\vspace*{0.1cm}
\renewcommand{\arraystretch}{1.0}
\centering
\begin{tabular}{|c@{\hskip .66cm}c@{\hskip .66cm}c|}
\hline
\hline
 \multicolumn{3}{c}{
Minimal Left Right Symmetric Model }\\
\hline
Matter & Generations &  $SU(2)_L \times SU(2)_R \times U(1)_{B-L} \times  SU(3)_C $\\
\hline
Fermions & & \\
$L_L$ & 3 & ($\mathbf{2}$, $\mathbf{1}$, $-1 $, $\mathbf{1}$) \\
$L_R$ & 3 & ($\mathbf{1}$, $\mathbf{2}$, $-1$, $\mathbf{1}$) \\
$Q_L$ & 3 & ($\mathbf{2}$, $\mathbf{1}$, $+\frac{1}{3}$, $\mathbf{3}$) \\
$Q_R$ & 3 & ($\mathbf{1}$, $\mathbf{2}$, $+\frac{1}{3}$, $\mathbf{3}$) \\
Scalars &   &  \\
$\Phi$ & 1  & ($\mathbf{2}$, $\overline{\mathbf{2}}$, $0$, $\mathbf{1}$) \\
$T_R$ & 1 & ($\mathbf{1}$, $\mathbf{3}$, $+2$, $\mathbf{1}$) \\
$T_L$ & 1  & ($\mathbf{3}$, $\mathbf{1}$, $+2$,  $\mathbf{1}$) \\
\hline
\hline
 \multicolumn{3}{c}{DM Candidates}\\
 \hline
Fermion  &  & \\
$\Psi$ &  1  & ($\mathbf{2}$, $\mathbf{2}$, $0$, $\mathbf{1}$) \\[.1cm] \hline
\hline
\hline
\multicolumn{3}{c}{Additional Matter for conserving D parity}\\
 \hline
Fermions  &  & \\
$F_c^\prime$ &  4  & ($\mathbf{1}$, $\mathbf{1}$, $0$ $\mathbf{3}$)$_L$ $\oplus$ ($\mathbf{1}$, $\mathbf{1}$, $0$, $\mathbf{3}$)$_R$ \\[.1cm] \hline
$F_d$ &  1  & ($\mathbf{2}$, $\mathbf{2}$, $0$ $\mathbf{1}$)$_L$ $\oplus$ ($\mathbf{2}$, $\mathbf{2}$, $0$, $\mathbf{1}$)$_R$ \\[.1cm] \hline
\hline
\multicolumn{3}{c}{Additional Matter for \cancel{D} parity}\\
 \hline
Fermions  &  & \\
$F_a$ &  4  & ($\mathbf{2}$, $\mathbf{1}$, $0$ $\mathbf{3}$)\\[.1cm] \hline
$F_b$ &  1  & ($\mathbf{1}$, $\mathbf{2}$, $0$, $\mathbf{3}$) \\[.1cm] \hline
$F_c$ &  4  & ($\mathbf{1}$, $\mathbf{1}$, $0$, $\mathbf{3}$) \\[.1cm] \hline
\hline
\end{tabular}
\end{table}
\paragraph{Beta functions}
The beta functions of the gauge couplings at two loops determine to a pretty good accuracy the unification and breaking of the $SU(2)_R$ scales. At two loops, the gauge couplings are only affected by their own running and the running of the top Yukawa coupling and hence one can consider the running of these quantities, without making any assumptions about the running of all of the other parameters of the theory. Obviously, for a specific model one must perform a complete running of all the parameters, but $\MR$ and $\MG$ can be determined to a great accuracy.
For the beta functions of the gauge couplings of the  models we present, we adopt the convention
\bea
\frac{d g_i}{dt} = \frac{b_i}{16\pi^2} g_i^3 +\frac{g_i^3}{(16 \pi^2)^2} \left[\sum_{j=1}^{j=3}  b_{ij} g_j^2 - c_i y_t^2\right],
\eea
where $g_i$ are the gauge couplings of each $i$ factor and $y_t$ is the gauge couplings of top Yukawa coupling. The coefficients $b_i$, $b_{ij}$ and $c_i$ are determined by group invariants and the matter content of the theory. The beta functions of the basic LR symmetric model are well known for the D parity conserving case, e.g. at one loop \cite{Lindner:1996tf} and at two loops \cite{Chakrabortty:2017mgi,Chakrabortty:2019fov} and can be easily checked 
with pyr@te \cite{pyr@te}. For the one loop beta functions of other models we use \cite{Lindner:1996tf} and pyr@te \cite{pyr@te} for cross checking and two loops computations. The coefficients of the beta functions for all the models we present in Section~\ref{sec:high_scale} are given in \Tabref{tab:beta_functions}.
 \begin{table}[]
     \centering
     \caption{Beta functions of the models we consider. The order of the indices is $i=L, R, BL, 3$.   \label{tab:beta_functions}}
     \vspace*{2mm}
     \begin{tabular}{|l|c|c|}
     \hline
     \hline
Model        & $b_i^{(1)}$  & $b_{ij}^{(2)}$ \\
\hline
LRSMD + $\Psi$  & $\left(
\begin{array}{c}
-5/3\\
-5/3\\
7\\
-7\\
\end{array}
\right)$    & 
$\left(\begin{array}{cccc}
209/6 & 9/2  &  27/2  &  12  \\  
9/2  &  209/6  &  27/2 &12  \\
81/2  &   81/2  &  115/2 & 4  \\  
9/2  &  9/2  &  1/2  &  -26  
\end{array}\right)$
\\
\hline
 LRSMD + $\Psi$ + $4\, F_c^\prime + $   $ \, F_d$ & $\left(
\begin{array}{c}
-1/3\\
-1/3\\
7\\
-13/3\\
\end{array}
\right)$   &
$\left(
\begin{array}{cccc}
307/6  & 15/2  & 27/2  & 12  \\
15 /2  & 307/6  & 27/2  & 12  \\
81 /2  & 81 /2  & 115/3  & 2  \\
9 /2  & 9 /2  & 1 /2  & 74/3  
\end{array}
\right)$
\\ 
\hline
LRSM\cancel{D} + $\Psi$  & $\left(
\begin{array}{c}
-7/3\\
-5/3\\
11/2\\
-7\\
\end{array}
\right)$  & 
$\left(
\begin{array}{cccc}
97/6  &  9/2  &  3/2 & 12   \\ 
9/2  &  209/6 & 27/2 & 12   \\
  9/2  &  81/2  &  61/2 & 4 \\
 9/2  &  9/2  &  1/2  &  -26  
 \end{array}\right)$\\
 \hline
 LRSM\cancel{D} + $\Psi +\, 4\, F_a + \, F_b + \, 4\, F_c$ &  
 $\left(
\begin{array}{c}
5/3\\
-2/3\\
11/2\\
-7/3\\
\end{array}
\right)$ &  
$\left(\begin{array}{cccc}
391/6  &  9/2   &  3/2  &  28  \\ 
9/2   & 565/12  & 27/2  &  16  \\ 
9/2  & 81/2  & 61/3  &  -1 \\  
21/2  & 6  &  1/2  &  188/3  
\end{array}
\right)$
\\
\hline
     \end{tabular}
 \end{table}
Due to GUT embedding the matching of the $LR$  gauge couplings to the SM gauge couplings  is given by:
\bea
\label{eq:matchingtoSM}
g^{ LR}_{B-L}(\MR) &=& z \sqrt{\left(\frac{2}{5\, z^2}  + \frac{3}{5}   \right)}\, g^{SM}_1(\MR),\nonumber\\
g^{LR}_{L}(\MR) &=& g^{SM}_{2}(\MR), \nonumber\\
g^{LR}_{C}(\MR) &=& g^{SM}_{3}(\MR),
\eea
where
\bea
\label{eq:matchingtoSMBLMR}
g^{LR}_{B-L}(\MR) = z \, g^{LR}_{R}(\MR).
\eea
Here $z$ is a real number of $\mathcal{O}(1)$ that is determined through the constraint of unification at $\MG$.
The gauge couplings obtained in the plots in \Figref{fig:running_models} correspond to the last running of the program that looks for convergence of the up and down running of the gauge couplings with the boundary conditions at $M_{EW}$ and gauge unification at $M_{\rm GUT}$. The matching to the SM model is given by:
\bea
\frac{1}{g^{2}_{1}(\MR)}&=& \frac{2}{5} \frac{1}{g^2_{B-L} (\MR)} + \frac{3}{5} \frac{1}{g^2_R(\MR)},\\
g^{SM}_2(\MR)&=&g^{LR}_L(\MR),\\
g^{SM}_3(\MR)&=&g^{LR}_3(\MR).
\eea
We have used the input parameters given in Table 1 of \cite{Chun:2021brv}.

\paragraph{Understanding the Need of Additional Matter}
In order to understand why the addition of color triplets can drastically lower the scale $\MR$, we consider the one loop beta function coefficients of the Minimal LRSM models (both D conserving and D breaking). These coefficients can be written as:
\bea
\label{eq:1loopbLRSM}
\left(
\begin{array}{c}
b_L\\
b_R\\
b_{B-L}\\
b_3
\end{array}\right)= 
\left(
\begin{array}{c}
-6\\ 
-6 \\ 
4/3\\
-29/3
\end{array}\right)+ N_{HBD} \left(
\begin{array}{c}
1/3\\ 
1/3 \\ 
0\\
0
\end{array}
\right) +
N_{H T_L} \left(
\begin{array}{c}
0\\ 
2/3 \\ 
3\\
0
\end{array}
\right)
+
N_{H T_L} \left(
\begin{array}{c}
2/3\\ 
0\\ 
3\\
0
\end{array}
\right)\!,
\eea
where $N_{HBD}$ is the number of Higgs bi-doublets, $N_{H T_L}$ and $N_{H T_R}$ is the number of Higgs Left-Triplets and Right-Triplets respectively. The first term on the right--hand side of~\eq{eq:1loopbLRSM} corresponds to the one-loop beta functions from gauge interactions and the three families of fermions of the LRSM. In the D parity conserving case both the left- and right-handed triplets $T_L$ and $T_R$ are present and $b_L=b_R=-5/3$. On the other hand, when only $T_R$ is included D parity is broken and $b_R> b_L=-7/2$. 

In order to obtain a phenomenologically viable scenario at low energy one needs to push the $M_R$ and $M_{\rm GUT}$ scales further apart compared to the minimal cases LRSMD + $\Psi$ and LRSM$\cancel{\rm D}$ + $\Psi$.
In \Figref{fig:running_models} $b_i<0$ implies a positive slope for $1/\alpha_i$. 
Indeed, the addition of color triplets makes $b_3$ less negative flattening the running of $1/\alpha_3$ and shifting $M_{\rm GUT}$ to higher scales and $M_R$ to lower scales.  However adding only color triplets is not enough because multiplets of all the other groups are required in order to ensure that $g_L$ and $g_R$ unify at high scale. In  particular, in order to increase the $g_R/g_L$ ratio more multiplets of $SU(2)_L$ than of $SU(2)_R$ are needed in order to keep the one loop beta function $b_R$ as negative as possible so that $g_R(\MR) > g_L (\MR)$  already at the scale $\MR$ (moreover, to overcome the running of $g_L$ the hierarchy between the two couplings needs to be strong enough).

\addcontentsline{toc}{section}{References}



\providecommand{\href}[2]{#2}\begingroup\raggedright\endgroup

\end{document}